\documentclass[10pt,journal,compsoc]{IEEEtran}

\usepackage[most]{tcolorbox}
\newcommand{\Beqref}[1]{Eq.~(\ref{#1})}
\newcommand{\Bfigref}[1]{Fig.~\ref{#1}}
\newcommand{\Bsecref}[1]{Sec.~\ref{#1}}
\newcommand{\Btabref}[1]{Tab.~\ref{#1}}

\newcommand{\x}[0]{\mathbf{x}}
\newcommand{\p}[0]{\mathbf{p}}
\newcommand{\y}[0]{\mathbf{y}}
\newcommand{\z}[0]{\mathbf{z}}

\newcommand{\dx}[0]{\;\mathrm{d}\x}
\newcommand{\dy}[0]{\;\mathrm{d}\y}
\newcommand{\dz}[0]{\;\mathrm{d}\z}

\newcommand{\abs}[1]{\lvert#1\rvert}

\newcommand{\Li}[0]{L_\text{i}}
\newcommand{\Lo}[0]{L_\text{o}}

\newcommand{\Le}[0]{L_\text{e}}
\newcommand{\fr}[0]{f_\text{r}}
\newcommand{\wi}[0]{\omega_\text{i}}
\newcommand{\wo}[0]{\omega_\text{o}}
\newcommand{\thetai}[0]{\theta_\text{i}}

\newcommand{\Ipx}[0]{I_\text{px}}
\newcommand{\Npx}[0]{N_\text{px}}

\newcommand{\est}[1]{\left\langle #1 \right\rangle}
\newcommand{\Expected}[1]{\mathbb{E}\left[#1\right]}
\newcommand{\Var}[1]{\mathbb{V}\left[#1\right]}
\newcommand{\estExpected}[1]{\tilde{\mathbb{E}}\left[#1\right]}
\newcommand{\estVar}[1]{\tilde{\mathbb{V}}\left[#1\right]}

\newcommand{\LossRRSNet}[0]{\mathcal{L}_{\text{RRSNet}}}

\usepackage{mathtools}

\let\oldparagraph\paragraph
\renewcommand{\paragraph}[1]{\oldparagraph*{\textbf{\textit{#1}}}}
\usepackage{enumitem} 
\usepackage{multicol} 
\usepackage{subcaption} 
\usepackage{adjustbox} 

\usepackage[ruled]{algorithm2e} 

\SetAlFnt{\small}
\SetAlCapFnt{\small}
\SetAlCapNameFnt{\small}
\SetAlCapHSkip{0pt}

\usepackage{booktabs,multirow}  
\usepackage{amsfonts}           
\usepackage{amsmath}            
\usepackage{amsthm}
\usepackage{siunitx}
\usepackage{colortbl}       
\usepackage{xcolor}
\usepackage{graphicx}
\usepackage{float}
\usepackage{subcaption}
\usepackage[numbers,sort&compress]{natbib}
\usepackage{soul}
\usepackage{makecell}
\usepackage{booktabs}

\usepackage{hyperref}
\usepackage[capitalise]{cleveref}

\begin{document}

\title{NRRS: Neural Russian Roulette and Splitting}




\author{Haojie Jin, Jierui Ren, Yisong Chen, Guoping Wang*, Sheng Li*,~\IEEEmembership{Member,~IEEE}
	\IEEEcompsocitemizethanks{
		\IEEEcompsocthanksitem Haojie Jin, Yisong Chen, Guoping Wang, and Sheng Li are with the School of Computer Science, Peking University, China.\\ E-mail: \
		\{jhj$|$chenyisong$|$wgp$|$lisheng\}@pku.edu.cn \\
		Guoping Wang and Sheng Li are also with the National Key Laboratory of Intelligent Parallel Technology.
		\IEEEcompsocthanksitem Jierui Ren is with the College of Future Technology, Peking University, China. \\
		E-mail: \{jerry@stu.pku.edu.cn\}
		\IEEEcompsocthanksitem Guoping Wang and Sheng Li are the corresponding authors.
	}
}

\IEEEtitleabstractindextext{
\begin{abstract}

	We propose a novel framework for Russian Roulette and Splitting (RRS) tailored to wavefront path tracing, a highly parallel rendering architecture that processes path states in batched, stage-wise execution for efficient GPU utilization. Traditional RRS methods, with unpredictable path counts, are fundamentally incompatible with wavefront's preallocated memory and scheduling requirements. To resolve this, we introduce a normalized RRS formulation with a bounded path count, enabling stable and memory-efficient execution.
	Furthermore, we pioneer the use of neural networks to learn RRS factors, presenting two models: NRRS and AID-NRRS. At a high level, both feature a carefully designed RRSNet that explicitly incorporates RRS normalization, with only subtle differences in their implementation. To balance computational cost and inference accuracy, we introduce Mix-Depth, a path-depth-aware mechanism that adaptively regulates neural evaluation, further improving efficiency.
	Extensive experiments demonstrate that our method outperforms traditional heuristics and recent RRS techniques in both rendering quality and performance across a variety of complex scenes.
\end{abstract}



}

\maketitle
\IEEEpeerreviewmaketitle

\section{Introduction}
\label{sec:introduction}



Path tracing is a widely adopted Monte Carlo method that approximates the rendering equation~\cite{Kajiya86} by stochastically tracing light transport paths from the camera into the scene. While path tracing is physically accurate, its convergence is slow due to high variance, especially in scenes with complex indirect lighting situations.
There are several approaches to improving the efficiency of unidirectional path tracing, including path guiding, multiple importance sampling, and Russian roulette and splitting (RRS)~\cite{ADRRS16}. In this paper, we primarily focus on RRS. RRS consists of two components: Russian roulette (RR) and splitting. RR aims to increase efficiency by terminating paths with low contributions, thereby reducing computational cost. Splitting seeks to improve efficiency by generating more paths with high contributions, which helps to reduce variance.

Traditional RRS methods are not suitable for wavefront path tracing. In the wavefront context, memory for work queues must be pre-allocated. If the path tracing algorithm does not perform splitting, it is sufficient to pre-allocate the work queues based on the number of pixels. However, after applying RRS, the maximum number of paths to be generated is unknown, making pre-allocation impossible. To address this issue, we propose a new RRS framework for wavefront path tracing. In short, we normalize the RRS factors and control the number of emitted paths within an upper bound, which guarantees theoretical convergence while maintaining practical tractability in wavefront-based scheduling.
Further details are provided in \Bsecref{sec:methods_rrs-framework}.


Under this new RRS framework, the assumptions of state-of-the-art RRS methods such as EARS~\cite{EARS22} are no longer valid. EARS employs separate evaluation branches for RR and splitting, but normalization may cross the boundary between them. Moreover, the normalization factor is not constant, which cannot be easily incorporated into its theoretical analysis. 

Recently, neural networks have demonstrated strong performance in various rendering tasks, such as neural radiance caching~\cite{NRC21} and path guiding~\cite{NPM23,NPG-NASG23}. Therefore, we leverage neural networks to learn the RRS factors.
Unlike previous methods, which cannot be easily adapted to our new framework, our approach allows for careful loss function design to ensure compatibility.
We propose two neural RRS methods, NRRS and AID-NRRS. To our knowledge, this is the first work to employ neural networks for estimating RRS factors. While neural networks introduce additional computational overhead, this cost may sometimes outweigh their benefits. To address this, and inspired by the common heuristic of disabling RR during the initial bounces, we introduce Mix-Depth (see \Bsecref{sec:methods_mix-depth}), which integrates other RRS strategies into NRRS and AID-NRRS to further improve efficiency.

To summarize, the key contributions of this paper are:

\begin{itemize}
	\item We propose a RRS framework for wavefront path tracing, making RRS both efficient and practical in the wavefront context.
	\item Building on this framework, we introduce two neural RRS models, NRRS and AID-NRRS. To the best of our knowledge, this is the first use of neural networks for estimating RRS factors with high efficiency.
	\item We present Mix-Depth, a general method for combining multiple RRS approaches to further enhance the efficiency of NRRS and AID-NRRS, called NRRS+ and AID-NRRS+ respectively. Mix-Depth is broadly applicable to any RRS method that shares the same cache structure, incurs minimal overhead, and is not limited to our neural approaches.
\end{itemize}
\section{Related Work}
\label{sec:related-work}

\subsection{Unidirectional Path Tracing}
Unidirectional path tracing (UPT) is a Monte Carlo-based method widely used for generating photorealistic images due to its simplicity. Rays are cast from the camera into the scene and continue to bounce within the environment until they either intersect a light source, exit the scene, or exceed a predefined bounce limit.

Since UPT is introduced by Kajiya~\cite{Kajiya86}, numerous unbiased improvements have been proposed to enhance its efficiency, including path guiding (PG)~\cite{PPG17,NPM23,NPG-NASG23}, RRS~\cite{ADRRS16,EARS22,MARS24}, control variates (CV)~\cite{NCV20,R-CV23}, and multiple importance sampling (MIS)~\cite{OMIS19,CMIS20,MMIS22}, among others. They orthogonally improve the efficiency of UPT. PG seeks to learn a better sampling distribution for generating new rays, improving the efficiency of Monte Carlo integration. MIS aims to optimally combine multiple sampling techniques to reduce variance. CV construct an auxiliary distribution such that subtracting it from the original yields a smoother, lower-variance residual that is easier to estimate. RRS adaptively allocate computational resources among paths to maximize overall efficiency. Our work focuses on RRS, but it can be easily combined with PG, CV, and MIS to further enhance efficiency.

\subsection{Russian Roulette and Splitting}
RRS consists of two techniques: Russian roulette (RR) and splitting. It was first introduced into path tracing in 1990~\cite{PTIS90}. RR terminates a path with a certain probability $1-p_{\text{rr}}$ ($0 < p_{\text{rr}} \leq 1$), allowing support for infinite path lengths. In contrast, splitting emits $n_{\text{s}}$ ($n_{\text{s}} \geq 1$) rays to estimate the local radiance. Typically, these probabilities are set heuristically. Throughput-based RR sets $p_{\text{rr}}$ equal to the current prefix path's throughput, which is the default setting in many renderers such as Mitsuba~\cite{mitsuba} and PBRT~\cite{pbrtv3}. \citet{GWTW05} analyzed the variance under a constraint on the total number of rays, and set $p_{\text{rr}}$ and $n_{\text{s}}$ according to both the prefix path and global properties. However, their approximation causes the method to degenerate into relying solely on the prefix path.

ADRRS~\cite{ADRRS16} combines RR and splitting into a unified framework. It determines the number of rays to generate based on a real-valued parameter $q$ ($0 \leq q < \infty$), which we refer to as the RRS factor in this paper. ADRRS also provides a method for computing $q$, defining it as the ratio of the estimated contribution of the entire path starting from the current prefix path to the estimated contribution of the pixel that the current prefix path contributes to. EARS~\cite{EARS22} employs a fixed-point iteration scheme to generate $q$ with the goal of maximizing efficiency, based on the theoretical analysis of \citet{EMMCRT97}. In EARS, $q$ does not distinguish between BSDF rays and NEE rays. MARS~\cite{MARS24} extends EARS to address this limitation. However, these RRS methods are all implemented in Megakernel-style path tracers and do not take the Wavefront architecture into consideration.

RRS has also been applied in other rendering algorithms, such as bidirectional path tracing (BDPT). \citet{Veach98} uses RR to reduce the number of visibility tests during the vertex connection step. \citet{ProxyBDPT24} employs RRS to maximize efficiency in reciprocal estimation.

\subsection{Neural Methods}
With the rapid development of deep learning, neural methods have been introduced into the context of path tracing. These approaches can be divided into two categories: \textit{Final Image Prediction} and \textit{Intermediate Quantity Prediction}. The former directly outputs the final image, while the latter estimates intermediate quantities such as radiance, hit point, or visibility during the path tracing process. Our method falls into the latter category.

\paragraph{Final Image Prediction}
One of the most popular directions is denoising~\cite{OptixDenoiser,OpenImageDenoise,WSKPD21,AFGSA21}, where a neural network produces a noise-free image from a noisy input and auxiliary buffers. As for Global illumination, \citet{CNNGI22} also use convolutional neural networks (CNNs) to predict single-bounce diffuse indirect illumination using direct illumination and auxiliary buffers as inputs.

\paragraph{Intermediate Quantity Prediction}
NRC~\cite{NRC21} uses a simple multi-layer perceptron (MLP) to predict local radiance. When certain conditions are met, the network is queried for local radiance instead of continuing path tracing.
Similarly, \citet{NIF23} use an MLP to predict visibility for calculating shadows from direct illumination. \citet{NBVH24} use an MLP to predict the intersection point and its normal; combined with scene-specific material parameters, this enables ray tracing to proceed using the network's predictions.
\section{Preliminary}
\label{sec:background}

\paragraph{Rendering Equation}
The core of solving the light transport problem is the rendering equation~\cite{Kajiya86} as:
\begin{equation}
	\label{eq:local_radiance_estimation}
	\Lo(\p, \wo) = \Le(\p, \wo)	+ \int_{\Omega} \Li(\p, \wi) \fr(\p, \wi, \wo) \abs{\cos\thetai} \;\mathrm{d}\wi \ ,
\end{equation}
which describes the relationship between the radiance $\Lo$ leaving a point $\p$ in direction $\wo$ and the incoming radiance $\Li$ arriving at $\p$ from direction $\wi$. Here, $\fr$ is the bidirectional scattering distribution function (BSDF) that describes how light is scattered at point $\p$, and $\thetai$ is the angle between the incoming direction $\wi$ and the normal vector at $\p$. $\Le(\p,\wo)$ denotes the radiance emitted from the point $\p$ in the direction $\wo$.

\paragraph{Monte Carlo Integration}
To solve the rendering equation, Monte Carlo integration is used to approximate the integral in \Beqref{eq:local_radiance_estimation}. The basic idea is to sample $N$ incoming directions $\wi$ according to a probability density function (PDF) $p(\wi\mid\wo, \p)$, and then estimate the integral by averaging the sampled values. Monte Carlo integration for the rendering equation is given by
\begin{equation}
	\label{eq:monte_carlo_integration}
	\langle \Lo(\p, \wo) \rangle = \Le(\p, \wo) + \frac{1}{N} \sum_{i=1}^{N} \frac{\Li(\p, \wi^{i}) \fr(\p, \wi^{i}, \wo) \abs{\cos\thetai^{i}} }{p(\wi^{i}\mid\wo,\p)}  .
\end{equation}

\paragraph{Unidirectional Path Tracing}
Unidirectional path tracing (UPT) solves the rendering equation by tracing paths from the camera to the light source. Each time a ray intersects a scene surface, Monte Carlo integration is performed to estimate the local radiance. This recursive process is repeated until the ray reaches a light source or a maximum number of bounces is reached.
Before performing Monte Car\-lo integration, RRS is used to determine how many samples should be taken.

\paragraph{Prefix and Suffix Paths}
We can divide the entire path $\z$ generated during the UPT process into two parts at the point $\p$ where Monte Carlo integration is performed. The first part is the prefix path $\x$, which extends from the camera to the point $\p$. The second part is the suffix path $\y$, which continues from $\p$ to the light source or infinity if the ray does not intersects with the scene. This decomposition allows the rendering equation to be reformulated as shown in \Beqref{eq:rendering_equation_prefix_suffix}. Here, $f$ represents the cumulative product of throughput terms (BSDF and cosine term as defined in \Beqref{eq:local_radiance_estimation}) along the path $\z$, while $g$ and $h$ denote the cumulative product of throughput terms along the prefix path $\x$ (excluding $\p$) and the suffix path $\y$ (including $\p$), respectively. $\mathcal{X}$ is defined as the set of prefix paths ending at the same point $\p$, and $\mathcal{Y}$ as the set of suffix paths starting from $\p$.

\begin{equation}	\label{eq:rendering_equation_prefix_suffix}
	\begin{aligned}
		I = \int f(\z)\dz = \int_{\mathcal{X}} g(\x)H(x) \dx \\
		\text{where } H(x) = \int_{\mathcal{Y}} h(\y) \dy.
	\end{aligned}
\end{equation}

\begin{figure}[ht]
	\centering
	\includegraphics[width=\linewidth]{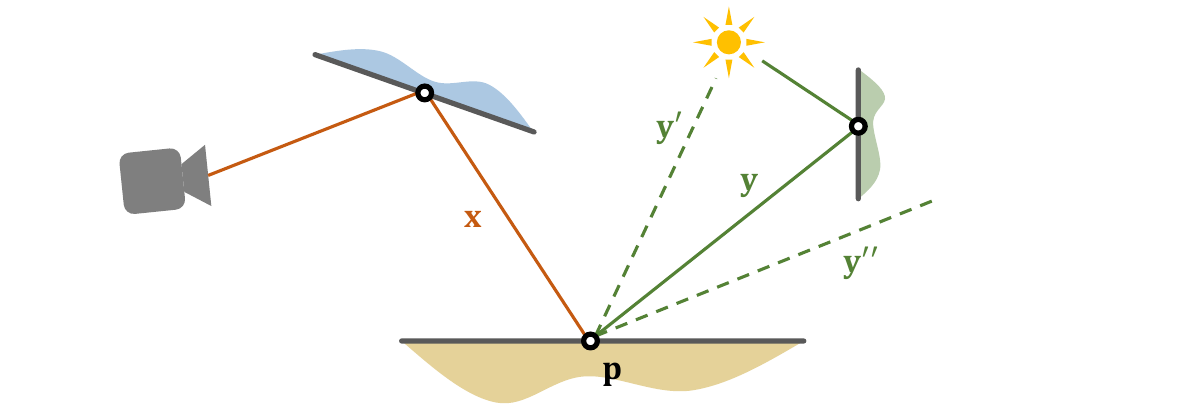}
	\caption{\textbf{Prefix and Suffix Paths.}
		The entire path $\z$ is divided into two parts at point $\p$. The prefix path $\x$ extends from the the camera to point $\p$. And the suffix path $\y$ continues from $\p$ to the light source or infinity.
	}
	\label{fig:path}
\end{figure}

\paragraph{Variance Analysis}
The variance analysis of RRS is provided by \citet{EMMCRT97}, and EARS~\cite{EARS22} adopts the same methodology. However, there are subtle differences between RR and splitting in how they affect variance. Here, we present only the final results; for detailed derivations, please refer to EARS~\cite{EARS22}.

For splitting, the relationship between the variance of the suffix path $\y$ and the variance of the entire path $\z$ is given by \Beqref{eq:variance_analysis_splitting}. Here, $n_{\text{s}}(\x)$ denotes the number of samples for suffix paths $\y$ (i.e., the splitting factor), and $p(\x)$ is the PDF of the prefix path $\x$. The derivative of $\Var{\langle I \rangle}$ with respect to $n_{\text{s}}(\x)$ is shown in \Beqref{eq:variance_analysis_splitting_derivative}.

\begin{equation}
	\label{eq:variance_analysis_splitting}
	\Var{\langle I \rangle} = \Var{\frac{g(\x)}{p(\x)}H(\x)} + \mathbb{E}\left[\left(\frac{g(\x)}{p(\x)}\right)^2\frac{\Var{\langle{H(\x)}\rangle}}{n_{\text{s}}(\x)}\right].
\end{equation}

\begin{equation}
	\label{eq:variance_analysis_splitting_derivative}
	\frac{\mathrm{d} \Var{\langle I \rangle}}{\mathrm{d} n_{\text{s}}(\x)} = -p(\x)\left(\frac{g(\x)}{p(\x)}\right)^2\frac{\Var{\langle{H(\x)}\rangle}}{n_{\text{s}}^2(\x)}.
\end{equation}

For RR, the relationship between the variance of the suffix path $\y$ and the variance of the whole path $\z$ is given by \Beqref{eq:variance_analysis_rr}, where $p_{\text{rr}}(\x)$ is the RR factor.

\begin{equation}
	\label{eq:variance_analysis_rr}
	\begin{aligned}
		\Var{\langle I \rangle} = & \quad \Var{\frac{g(\x)}{p(\x)}H(\x)}                                                                                                                        \\
		                          & + \mathbb{E}\left[\left(\frac{g(\x)}{p(\x)}\right)^2\left(\frac{\mathbb{E}\left[{\langle{H(\x)}\rangle}^2\right]}{p_{\text{rr}}(\x)}-H^2(\x)\right)\right].
	\end{aligned}
\end{equation}
\begin{equation}
	\label{eq:variance_analysis_rr_derivative}
	\frac{\mathrm{d} \Var{\langle I \rangle}}{\mathrm{d} p_{\text{rr}}(\x)} = -p(\x)\left(\frac{g(\x)}{p(\x)}\right)^2\frac{\mathbb{E}\left[{\langle{H(\x)}\rangle^2}\right]}{p_{\text{rr}}^2(\x)}.
\end{equation}

\paragraph{Unified RRS Framework}
ADRRS provides a unified framework for RRS. Given an RRS factor $q$, the number of samples $N_q$ is determined as shown in \Beqref{eq:rrs_factor}. During the UPT process, the RRS factor $q$ specifies how many samples should be taken before applying Monte Carlo integration at each intersection point.
\begin{equation}
	\label{eq:rrs_factor}
	N_q =
	\begin{cases}
		\lfloor{q}\rfloor + 1 & \text{with the probability } q - \lfloor{q}\rfloor \\
		\lfloor{q}\rfloor     & \text{otherwise.}                                  \\
	\end{cases}
\end{equation}

\begin{figure}[htbp]
	\centering
	\includegraphics[width=\linewidth]{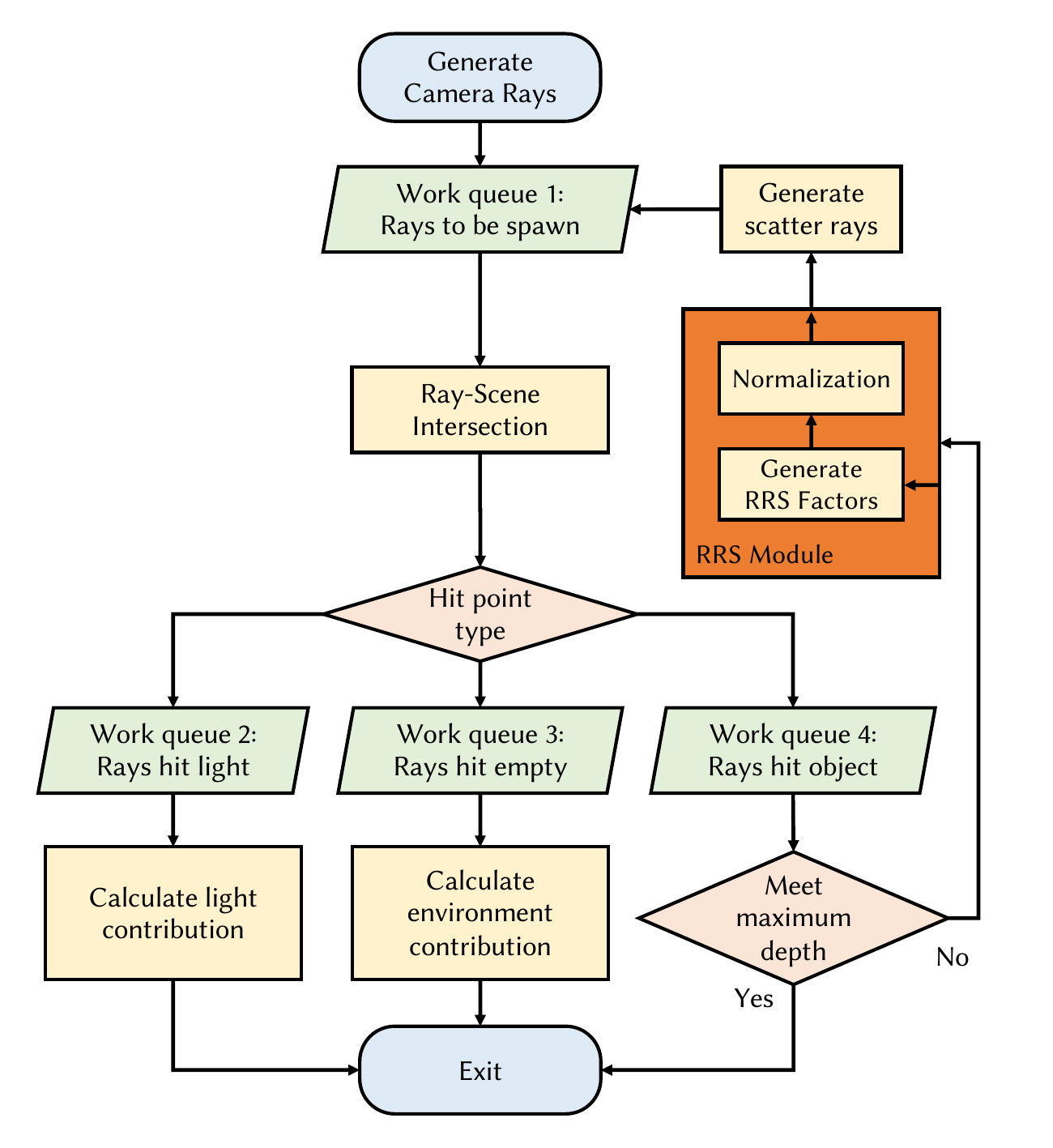}
	\caption{\textbf{Wavefront Architecture with RRS module.}
        The essence of the wavefront path tracing architecture lies in decomposing the control flow of path tracing into a staged dataflow. Leveraging a work-queue-based mechanism, it is expected to achieve highly parallel, low-divergence, and load-balanced execution.
Our RRS module regulates the generation and termination of paths, maintaining stable work queue sizes and providing a predictable workload scheduling mechanism for the GPU.
	}
	\label{fig:wavefront}
\end{figure}

\paragraph{Wavefront Architecture}
\citet{wavefront13} proposed the wavefront architecture to improve the efficiency of UPT. In contrast to the traditional megakernel architecture, which performs all operations for a ray within a single kernel, the wavefront architecture separates the computation of a ray into several stages, each implemented by a different kernel. This separation reduces thread divergence within each kernel and thus improves performance. The wavefront architecture is illustrated in \Bfigref{fig:wavefront}. 
In this framework, camera rays intersect the scene and are dispatched to four work queues. Rays hitting surfaces spawn new rays based on RRS factors, while those hitting lights or missing geometry terminate. The process repeats until the maximum path depth is reached.
In this example, next event estimation (NEE) is disabled, and a maximum path depth is enforced. 

Work queues must be pre-allocated under the wavefront architecture. In traditional UPT, the RRS module typically only performs RR, so the size of the work queues does not increase. However, once splitting is applied, the maximum number of paths to be generated becomes unknown, making pre-allocation infeasible. This unpredictability also leads to uncontrollable memory usage and scheduling challenges. To address this problem, we propose a new RRS framework, which will be discussed in \Bsecref{sec:methods_rrs-framework}. At a high level, we perform normalization at each depth to ensure that the number of rays remains within a specified limit.

\paragraph{Discussion on Neuralization of RRS}
A straightforward way to incorporate neural networks into RRS is to replace the local cache structure with a neural network.
For ADRRS~\cite{ADRRS16}, only the estimation of the mean of the outgoing radiance $\Lo(\x,\wo)$ is required, where $\x$ is the prefix path ending at point $\p$ and $\wo$ is the outgoing direction. However, ADRRS only considers the expected contribution of suffix paths $\z$ starting from $\p$, without accounting for their variance or computational cost.
For EARS~\cite{EARS22}, it needs to estimate both the mean and variance of $\Lo(\x,\wo)$, as well as the expected computational cost of tracing suffix paths $\z$. It is challenging to accurately estimate all three quantities simultaneously using a single simple neural network, as discussed in detail in \Bsecref{sec:results_ablation_earsnn}.
Moreover, both ADRRS and EARS cannot be easily adapted to our normalization architecture by simply modifying their theoretical formulations. In particular, for EARS, the normalization process undermines its theoretical foundation, making it no longer optimal.
In the following section, we introduce a neural network-based RRS method designed to address this issue.

\section{Methods}
\label{sec:methods}
In this section, we first propose a theoretical RRS framework for wavefront UPT. Next, we introduce two neural RRS methods, called NRRS and AID-NRRS, specifically designed to operate within this framework. Finally, we present Mix-Depth, a method for combining multiple RRS techniques to further improve efficiency.

\subsection{Normalization RRS framework for Wavefront}
\label{sec:methods_rrs-framework}

\subsubsection{Problem Statement}
Traditional RRS algorithms generate a stochastic number of secondary samples determined by per-path factors $q_{\text{orig}}$.  
While this flexibility is acceptable in recursive path tracing, it becomes problematic in wavefront architectures, which process path states in large, preallocated batches on the GPU.  
Since wavefront path tracing relies on a fixed-size workqueue and synchronized execution, an uncontrolled number of samples causes severe instability, memory overflow, and unpredictable scheduling.  
A simple truncation of excessive samples, however, introduces statistical bias, particularly when a valid contribution is discarded entirely.

\subsubsection{Normalization Process}
To resolve this structural incompatibility, we introduce a normalization step that globally constrains the total number of samples.  
Hence, we define the normalization operation as:
\begin{equation}
q_{\text{norm}}^{(i)} = F_{\text{norm}} \cdot q_{\text{orig}}^{(i)},
\label{eq:q_norm}
\end{equation}
where the normalization factor $F_{\text{norm}}$ is computed dynamically as
\begin{equation}
F_{\text{norm}} = \frac{\Npx}{\sum_i q_{\text{orig}}^{(i)}} .
\label{eq:F_norm}
\end{equation}
This guarantees that
\begin{equation}
\sum_i q_{\text{norm}}^{(i)} = \Npx,
\end{equation}
where $\Npx$ denotes the total number of pixels (or equivalently, the preallocated size of the primary workqueue).  
Intuitively, $F_{\text{norm}}$ acts as a global scaling factor enforcing a conservation constraint on sample count, ensuring that the expected total number of scattered rays matches the rendering budget per iteration.

\subsubsection{Boundedness and Architectural Stability}
The normalization process guarantees that the expected number of generated samples equals $\Npx$.  
Hence, the algorithm achieves bounded sample growth, allowing stable GPU execution and predictable memory allocation within the wavefront pipeline.
\label{subsec:bounded_stable}

\paragraph{Statement}
Under the normalization defined in \Beqref{eq:q_norm}--\Beqref{eq:F_norm},
let $X_i$ denote the random number of scattered samples generated from
the $i$-th parent path in one wavefront iteration, and let
$\mathcal{I}$ be the index set of active parent paths.
The total number of generated samples is
\[
S = \sum_{i\in\mathcal{I}} X_i .
\]
Then, the expected total number of samples is bounded as
\[
\mathbb{E}[S] = \Npx,
\]
and the conditional variance admits the upper bounds
\[
\mathrm{Var}(S \mid \{q_{\text{norm}}^{(i)}\}) \le \rho \le \Npx,
\qquad
\mathrm{Var}(S \mid \{q_{\text{norm}}^{(i)}\}) \le \tfrac{1}{4}|\mathcal{I}|,
\]
where $\rho = \sum_{i\in\mathcal{I}} r_i$ and
$q_{\text{norm}}^{(i)} = k_i + r_i$ with
$k_i=\lfloor q_{\text{norm}}^{(i)}\rfloor\in\mathbb{N}$ and $r_i\in[0,1)$.

\paragraph{Proof}
\textbf{(1) Modeling and single-path statistics.}
Each normalized RRS factor is decomposed as
\[
q_{\text{norm}}^{(i)} = k_i + r_i,\quad r_i\in[0,1),
\]
and the number of generated children is sampled via
the ``integer-splitting + residual Bernoulli'' rule:
\[
X_i = k_i + B_i, \qquad B_i \sim \mathrm{Bernoulli}(r_i),
\]
with $\{B_i\}$ conditionally independent given $\{q_{\text{norm}}^{(i)}\}$.
Then,
\begin{align}
\mathbb{E}[X_i \mid q_{\text{norm}}^{(i)}] = k_i + r_i = q_{\text{norm}}^{(i)}, \\ 
\mathrm{Var}(X_i \mid q_{\text{norm}}^{(i)}) = r_i(1-r_i) \le \tfrac{1}{4}.
\end{align}

\textbf{(2) Expectation boundedness.}
Using the normalization
$q_{\text{norm}}^{(i)} = F_{\text{norm}} q_{\text{orig}}^{(i)}$
with
$F_{\text{norm}} = \frac{\Npx}{\sum_i q_{\text{orig}}^{(i)}}$,
we have
\[
\mathbb{E}[S \mid \{q_{\text{orig}}^{(i)}\}]
= \sum_i \mathbb{E}[X_i \mid q_{\text{norm}}^{(i)}]
= \sum_i q_{\text{norm}}^{(i)}
= \Npx.
\]
Taking total expectation yields
\[
\boxed{\mathbb{E}[S] = \Npx.}
\]
Thus, the expected number of generated samples is globally bounded
and independent of scene complexity.

\textbf{(3) Variance upper bound.}
Under conditional independence,
\begin{align}
\mathrm{Var}(S \mid \{q_{\text{norm}}^{(i)}\})
= \sum_i \mathrm{Var}(X_i \mid q_{\text{norm}}^{(i)}) \\
= \sum_i r_i(1-r_i)
\le \sum_i r_i
= \rho.
\end{align}
Because $\sum_i q_{\text{norm}}^{(i)} = \sum_i k_i + \rho = \Npx$,
we obtain $\rho \le \Npx$, leading to
\[
\boxed{\mathrm{Var}(S \mid \{q_{\text{norm}}^{(i)}\}) \le \rho \le \Npx.}
\]
Using $r_i(1-r_i)\le \tfrac{1}{4}$ further gives a universal bound
independent of scene statistics:
\[
\boxed{\mathrm{Var}(S \mid \{q_{\text{norm}}^{(i)}\}) \le \tfrac{1}{4}|\mathcal{I}|.}
\]
\hfill $\square$

\paragraph{Discussion}
The above results indicate that:
(1) the expected sample count is strictly bounded by the pixel number $\Npx$,
and (2) its variance grows linearly with the number of active paths.
Consequently, workqueues can be safely preallocated with fixed size $\Npx$,
ensuring stable GPU memory usage and predictable wavefront scheduling.
Normalization thus provides the mathematical foundation for architectural stability.

\subsubsection{Rate-Control Factor}

\label{subsec:rate_control}

Although normalization ensures that the expected number of generated samples satisfies $\mathbb{E}[S]=\Npx$, the stochastic nature of RRS can still cause the realized count $S$ to slightly exceed $\Npx$ with a small probability. To further suppress this overflow probability, we introduce a mildly scaled \emph{rate-control factor} $f_{\text{rate}}\in(0,1)$ that adjusts the normalized RRS factors.

\paragraph{Definition}
Given the normalized factors $q_{\text{norm}}^{(i)}$, define the rate-controlled factors as
\begin{equation}
q_{\text{norm}}'^{(i)} = f_{\text{rate}}\cdot q_{\text{norm}}^{(i)} .
\label{eq:q_norm_rate}
\end{equation}
The expected number of generated samples becomes
\begin{equation}
\mathbb{E}[S'] = f_{\text{rate}}\Npx,
\end{equation}
where $S'$ denotes the total number of scattered samples after rate control.

\paragraph{Overflow probability}
By Bernstein’s inequality, the probability that $S'$ exceeds the queue capacity $\Npx$ is exponentially small:
\begin{equation}
\mathbb{P}(S' \ge \Npx)
\;\le\;
\exp\!\left(
-\frac{(1-f_{\text{rate}})^2\,\Npx}{\,2f_{\text{rate}}+\tfrac{2}{3}(1-f_{\text{rate}})\,}
\right).
\label{eq:bernstein_frate}
\end{equation}
For instance, with $f_{\text{rate}}=0.85$, the exponent is approximately $-0.0125\,\Npx$. We provide the detailed derivation in the supplemental material.

\paragraph{Adaptive variant}
In practice, we also use an adaptive form $f_{\text{rate}}' = \alpha f_{\text{rate}}$ to allow finer control, where $\alpha\in(0,1]$ is reduced upon detecting overflow:
\begin{equation}
\alpha \leftarrow \alpha \cdot (1 - \epsilon), \qquad \epsilon \approx 0.01.
\label{eq:f_rate_prime}
\end{equation}

The rate-control factor provides a safety margin atop normalization: while normalization guarantees the \emph{expected} sample count is bounded, rate control ensures the \emph{realized} count rarely exceeds the preallocated capacity. Together they deliver both theoretical and practical stability for RRS in wavefront path tracing.

\subsubsection{Extended Discussion}

Normalization serves as a critical bridge between the theoretical design of RRS and the practical constraints of wavefront path tracing. 
Architecturally, it transforms an unbounded stochastic sampling process into a bounded, resource-aware one, ensuring predictable GPU scheduling and stable memory utilization. 
Statistically, normalization enforces a global budget constraint on sample generation, preserving expectation consistency and mitigating the catastrophic bias that would otherwise result from naive sample truncation.

Nevertheless, normalization also introduces theoretical challenges. 
Because $F_{\text{norm}}$ is dynamic and scene-dependent, the normalized factors $q_{\text{norm}}$ are no longer linearly proportional to their original counterparts $q_{\text{orig}}$, potentially inverting the behavior of Russian Roulette and splitting. 
This violates the proportional-invariance assumption underlying existing analytical RRS schemes such as EARS, thereby compromising their theoretical consistency. 
Hence, while normalization is indispensable for boundedness and stability, it simultaneously demands a reformulation of RRS that remains coherent under dynamic scaling.

To address this limitation, we introduce a learning-based paradigm in which a neural network directly models the normalized RRS process. 
Unlike handcrafted heuristics that rely on static probabilistic formulations, our proposed neural RRS (NRRS and AID-NRRS) learns a data-driven mapping from local path statistics and scene attributes to normalized RRS factors, naturally adapting to the stochastic constraints imposed by $F_{\text{norm}}$. 
This neural formulation implicitly maintains statistical coherence under normalization, while preserving the architectural efficiency and memory stability of the wavefront framework. 
In essence, the neural approach provides a principled and flexible pathway to integrate normalization into RRS, achieving both theoretical rigor and practical scalability within a unified framework.

\begin{figure*}[th]
	\centering
   \begin{subfigure}{0.5\textwidth}
       \centering
       \includegraphics[scale=0.43]{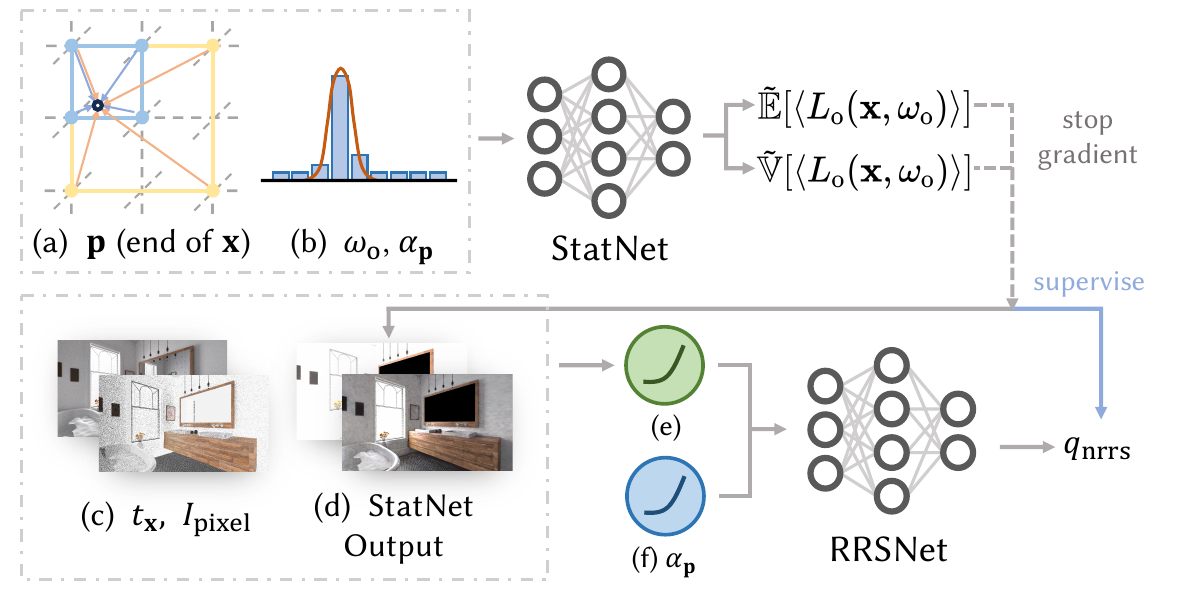}
       \caption{Schematic of NRRS}
       \label{fig:nrrs}
   \end{subfigure}%
   \begin{subfigure}{0.5\textwidth}
       \centering
       \includegraphics[scale=0.43]{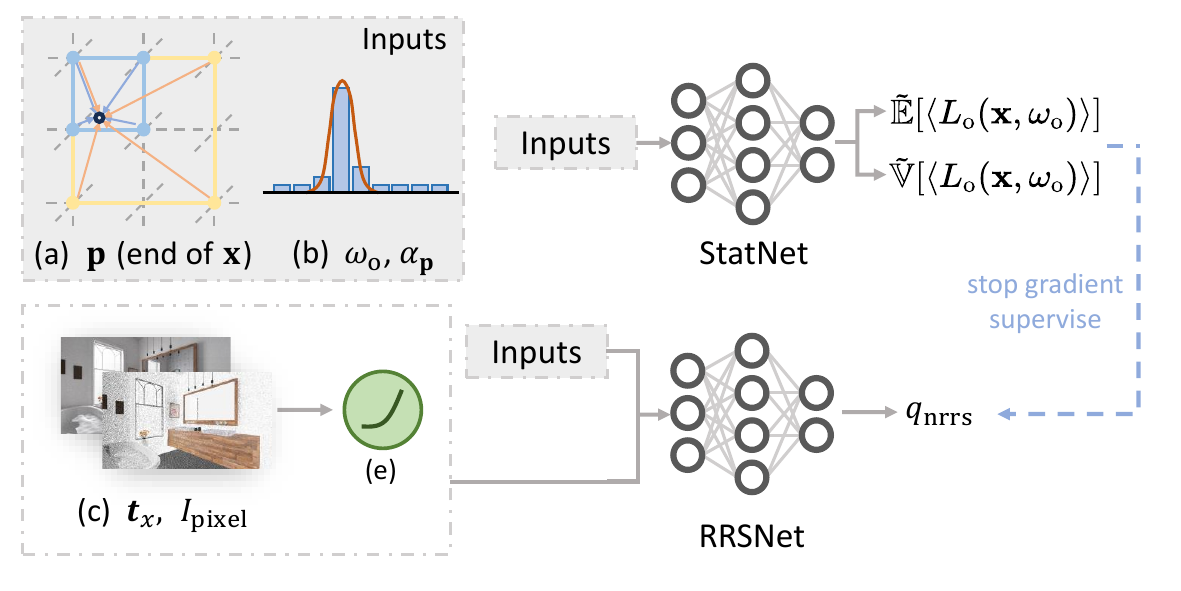}
       \caption{Schematic of AID-NRRS}
       \label{fig:aid-nrrs}
   \end{subfigure}
   	\caption{
		\textbf{(A)} NRRS consists of two neural networks: \textbf{StatNet} and \textbf{RRSNet}. StatNet predicts the local radiance statistics $\Lo(\mathbf{x}, \omega_o)$, while RRSNet predicts the RRS factors $q_{\text{nrrs}}$. StatNet takes as input the position $\mathbf{p}$, the outgoing direction $\omega_o$, and the surface roughness $\alpha_{\mathbf{p}}$, and outputs both the mean and variance of the local radiance. These predicted statistics serve both as direct inputs to RRSNet (d) and as supervisory signals for training its output. (a) The position $\mathbf{p}$ is encoded using a learnable multi-resolution hash grid~\cite{TCNN}. (b) The direction $\omega_o$ and roughness $\alpha_{\mathbf{p}}$ are encoded using one-blob encoding~\cite{NIS19}. The predicted radiance statistics (after Box–Cox transformation) (e), together with throughput $t_{\mathbf{x}}$, pixel radiance $I_{\text{pixel}}$, and remapped roughness (f), are fed into RRSNet to predict the resampling factor $q_{\text{nrrs}}$ for point $\mathbf{p}$.
        \textbf{(B)} AID-NRRS is similar to NRRS, except that RRSNet omits the StatNet outputs and directly takes the position $\p$ and outgoing direction $\wo$ as inputs. Note that StatNet and RRSNet each have their own Inputs module with separate trainable parameters.
	}
	\label{fig:nrrs-network}   
\end{figure*}

\subsection{Neural RRS}

With the success of neural methods~\cite{NRC21,NPM23,NBVH24} in rendering, we further employ neural networks to predict RRS factors, leveraging their ability to approximate target functions continuously and more accurately than discrete heuristics. Furthermore, by designing appropriate loss functions (see \Bsecref{sec:methods_loss-design}), we can mitigate issues arising from the behavioral changes of RRS factors before and after normalization. In other words, the side effect of the normalization process can be explicitly incorporated into the loss design. Note that our neural RRS methods are scene-specific.

\subsubsection{Network Structure}
\label{sec:methods_network-structure}
We refer to our neural RRS methods as NRRS and AID-NRRS, which differ only in the design of RRSNet (see \Bfigref{fig:nrrs-network}). Both methods share the same training strategy.

The main challenge in NRRS and AID-NRRS is that the ground truth for RRS factors yielding optimal efficiency is unknown, making supervised training infeasible. Fortunately, under our new RRS framework, the cost of rendering each frame is nearly constant, as normalization is performed at every bounce. With this fixed cost, maximizing efficiency can be reformulated as minimizing the variance of pixel radiance. Since the relationship between pixel variance and the RRS factors is known from \Beqref{eq:variance_analysis_splitting_derivative} and \Beqref{eq:variance_analysis_rr_derivative}, we can design a loss function to minimize pixel variance and propagate gradients to the RRS factors accordingly.

As shown in \Beqref{eq:variance_analysis_splitting_derivative} and \Beqref{eq:variance_analysis_rr_derivative}, we need certain statistics of the local radiance $\Lo(\x,\wo)$ to enable gradient transfer. Therefore, we design our network to consist of two components: StatNet and RRSNet. The details of these two networks are as follows:

\paragraph{StatNet}
StatNet predicts the local radiance $\Expected{\est{\Lo(\x,\wo)}}$ and its variance $\Var{\est{\Lo(\x,\wo)}}$, both of which are essential for propagating gradients from pixel variance to the RRS factors. Inspired by NRC~\cite{NRC21}, we design StatNet to take as input the outgoing direction $\wo$, the position $\p$, and auxiliary information such as the roughness at $\p$.

\paragraph{RRSNet}
RRSNet is responsible for predicting the RRS factors, \- $q_{\text{nrrs}}(\x)$. We have two design choices: (1) using the outputs of StatNet as inputs and supervisory signals or (2) using the outputs of StatNet only as supervisory signals. We refer to the first design as NRRS (see \Bfigref{fig:nrrs} and the second as AID-NRRS (see \Bfigref{fig:aid-nrrs}).

For NRRS (\Bfigref{fig:nrrs}), we provide RRSNet with the following necessary information:
\begin{itemize}
	\item[(1)] the prefix path $\x$, specifically its throughput $t_{\x}$.
	\item[(2)] the endpoint $\p$, specifically its roughness $\alpha_{\p}$.
	\item[(3)] the full path, specifically the pixel value $I_{\text{pixel}}$ to which $\x$ contributes.
	\item[(4)] the suffix path $\y$, specifically the local radiance statistics estimations $\estExpected{\est{\Lo(\x,\wo)}}$, $\estVar{\est{\Lo(\x,\wo)}}$ from StatNet.
\end{itemize}

For AID-NRRS (\Bfigref{fig:aid-nrrs}), the first three items are the same as NRRS. As for the last item, we directly use the position $\p$ and the outgoing direction $\wo$ instead of the local radiance statistics. Note that StatNet and RRSNet have its own Inputs module (trainable parameters) respectively.

For NRRS, we do not use $\p$ or $\wo$ as inputs, so that RRSNet learns the relationship between the input features and the RRS factors, rather than memorizing the distribution of RRS factors across the scene. This design choice improves the generalization ability of RRSNet and makes the learning process more efficient. However, it requires StatNet to be executed during inference (unlike AID-NRRS), which introduces additional overhead. Therefore, NRRS is more suitable for scenes where the RRS factors change frequently throughout the scene.


The Loss Design (\Bsecref{sec:methods_loss-design}) and Mix-Depth (\Bsecref{sec:methods_mix-depth}) is the same for both NRRS and AID-NRRS, so we use NRRS as an example in the following subsections.

\subsubsection{Loss Design}
\label{sec:methods_loss-design}
For StatNet, we use the local radiance estimation $\est{\Lo(\x,\wo)}$ obtained during the path tracing process as ground truth. StatNet is trained to predict the average local radiance $\Expected{\est{\Lo(\x,\wo)}}$ and its statistic $\Var{\est{\Lo(\x,\wo)}}$. We use relative MSE~\cite{relativeL2} as our loss function, as shown in \Beqref{eq:loss-function_StatNet}. Here, $\estExpected{\est{\Lo(\x,\wo)}}$ and $\estVar{\est{\Lo(\x,\wo)}}$ are the outputs of StatNet, and $\text{sg}(\cdot)$ indicates that its argument is treated as a constant during optimization. As for $\Var{\Lo(\x,\wo)}$, we regard $\estExpected{\est{\Lo(\x,\wo)}}$ as mean value and calculate one-sample variance of $\est{\Lo(\x,\wo)}$ as the ground truth.

\begin{equation}
	\begin{aligned}
		\mathcal{L}_{\text{StatNet}} = & \quad \mathcal{L}_2(\estExpected{\est{\Lo(\x,\wo)}}, \est{\Lo(\x,\wo)})   \\
		                               & + \mathcal{L}_2(\estVar{\est{\Lo(\x,\wo)}}, \Var{\est{\Lo(\x,\wo)}}),     \\
		\text{where }                  & \mathcal{L}_2(a,b)= \frac{\left(a - b\right)^2}{\text{sg}(b)^2+\epsilon}. \\
	\end{aligned}
	\label{eq:loss-function_StatNet}
\end{equation}

\begin{figure}[htbp]
	\centering
	\includegraphics[width=\linewidth]{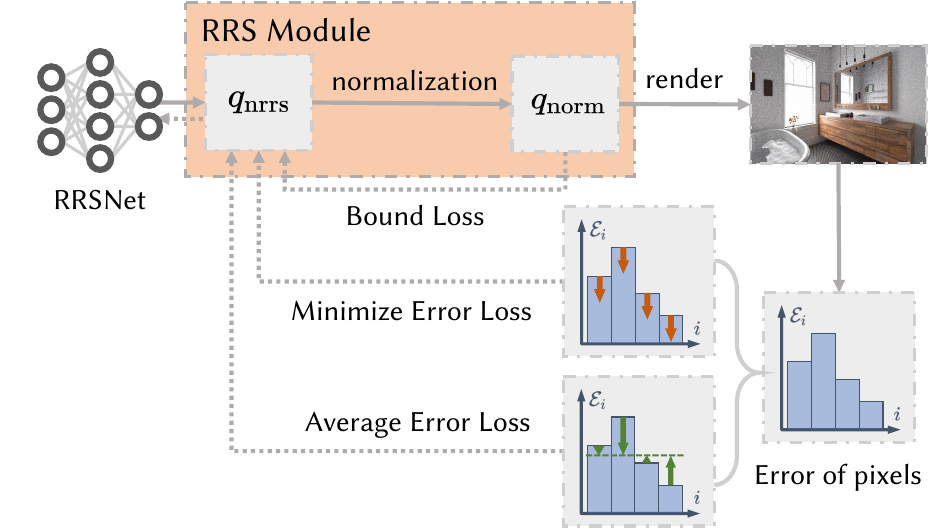}
	\caption{\textbf{Loss Function of RRSNet.}
	The loss function of RRSNet includes three components: Minimize Error Loss, Average Error Loss, and Bound Loss. 
    Solid arrows indicate the forward process—from RRS factor prediction and normalization to scattered path generation and pixel error computation. Dashed arrows represent the backward process, where gradients propagate through Minimize Error Loss and Average Error Loss, while Bound Loss directly constrains the original RRS factors.
	}
	\label{fig:loss}
\end{figure}

As for RRSNet, the loss function with three components is as:
\begin{equation}
	\LossRRSNet = \gamma_1\mathcal{L}_{\text{avg}} + \gamma_2 \mathcal{L}_{\text{min}} + \gamma_3 \mathcal{L}_{\text{rrs}}.
	\label{eq:loss-function_rrsnet}
\end{equation}
First, we explain how to transfer the gradient from the pixel error to the RRS factors. Then, we discuss each of the three loss functions in detail. \Bfigref{fig:loss} illustrates how the three components of the loss function interact and contribute to the overall optimization. The loss function of RRSNet consists of three components: Minimize Error Loss ($\mathcal{L}_{\text{min}}$), Average Error Loss ($\mathcal{L}_{\text{avg}}$), and Bound Loss ($\mathcal{L}_{\text{rrs}}$). Wherein, solid arrows represent the forward process: RRSNet predicts the original RRS factors $q$, which are then normalized to $q_{\text{norm}}$ by the RRS module. The normalized RRS factors are used to generate scattered paths, whose radiance contributions are accumulated to the corresponding pixel, and the error of pixels is computed. Dashed arrows indicate the backward process: gradients from the pixel error are propagated back to the RRS factors via the Minimize Error Loss and Average Error Loss, while the Bound Loss directly constrains the original RRS factors.

\paragraph{Gradient Transfer}
The relationship between the pixel variance and a specific prefix path $\x$ is defined by \Beqref{eq:variance_analysis_splitting_derivative} and \Beqref{eq:variance_analysis_rr_derivative}. Using these two equations, we derive how the pixel variance depends on the RRS factors under our network design.

We first consider RR. The input (denoted as $\theta$) of RRSNet can correspond to many different prefix paths $\x$, which we denote as the set $\Theta$. Therefore, the derivative of the pixel variance $\Var{\langle I \rangle}$ with respect to the RR factors $p_{\text{rr}}(\x)$ is the integral of $\frac{\mathrm{d} \Var{\langle I \rangle}}{\mathrm{d} p_{\text{rr}}(\x)}$ over the set $\Theta$, as shown in \Beqref{eq:gradient_transfer_rr}:
\begin{equation}
	\frac{\mathrm{d} \Var{\langle I \rangle}}{\mathrm{d} p_{\text{rr}}(\theta)} = \int_{\Theta} \frac{\mathrm{d} \Var{\langle I \rangle}}{\mathrm{d} p_{\text{rr}}(\x)}\,\mathrm{d}\x.
	\label{eq:gradient_transfer_rr}
\end{equation}
We use the Monte Carlo method to estimate this integral, collecting all necessary information during the UPT process. The one-sample Monte Carlo estimation is shown in \Beqref{eq:gradient_transfer_rr_mc}. Notably, the unstable term $p(\x)$ in \Beqref{eq:variance_analysis_rr_derivative} disappears, which makes the training process more stable.
\begin{equation}
	\est{\frac{\mathrm{d} \Var{\langle I \rangle}}{\mathrm{d} p_{\text{rr}}(\theta)}} = -\left(\frac{g(\x)}{p(\x)}\right)^2\frac{\mathbb{E}\left[{\langle{H(\x)}\rangle^2}\right]}{p_{\text{rr}}^2(\x)}.
	\label{eq:gradient_transfer_rr_mc}
\end{equation}

The deduction for the splitting case is similar. We can obtain the derivative of the pixel variance with respect to the splitting factors $n_{\text{s}}(\x)$ as shown in \Beqref{eq:gradient_transfer_splitting_mc}. Note that, as in EARS~\cite{EARS22} and MARS~\cite{MARS24}, we ignore the variance introduced by stochastic rounding.
\begin{equation}
	\est{\frac{\mathrm{d} \Var{\langle I \rangle}}{\mathrm{d} n_{\text{s}}(\theta)}} = -\left(\frac{g(\x)}{p(\x)}\right)^2\frac{\Var{\langle{H(\x)}\rangle}}{n_{\text{s}}^2(\x)}.
	\label{eq:gradient_transfer_splitting_mc}
\end{equation}

With the help of \Beqref{eq:gradient_transfer_rr_mc} and \Beqref{eq:gradient_transfer_splitting_mc}, we only need to design the loss function with respect to the pixel variance, and then the gradient can be propagated to the RRS factors for optimization.

\paragraph{Minimize Error Loss}
$\mathcal{L}_{\text{min}}$ is the loss function used to minimize the pixel error. We use the relative MSE to measure the pixel error. Because our method is unbiased, the MSE is equal to the variance. Here, $\epsilon=0.01$ is a small constant to maintain numerical stability.

\begin{equation}
	\mathcal{L}_{\text{min}} = \sum_{i=1}^{\Npx} \mathcal{E}_{i},~\text{where } \mathcal{E}_{i} = \frac{\Var{\langle I_{i} \rangle}}{I_{i}^2+\epsilon}.
	\label{eq:loss-function_min}
\end{equation}

\paragraph{Average Error Loss}
$\mathcal{L}_{\text{avg}}$ aims to make the error across pixels more uniform. This helps prevent too many rays from concentrating on a few pixels.

\begin{equation}
	\mathcal{L}_{\text{avg}} = \sum_{i=1}^{\Npx} \left(\mathcal{E}_i-\mathcal{E}_{\text{avg}}\right)^2,~\text{where } \mathcal{E}_{\text{avg}} = \sum_{i=1}^{\Npx} \frac{\mathcal{E}_{i}}{\Npx}.
	\label{eq:loss-function_avg}
\end{equation}

\paragraph{Bound Loss}
$\mathcal{L}_{\text{min}}$ will cause the RRS factors to grow toward infinity, as its gradient is always negative (see \Beqref{eq:gradient_transfer_rr_mc} and \Beqref{eq:gradient_transfer_splitting_mc}). Moreover, $\mathcal{L}_{\text{avg}}$ only fine-tunes the RRS factors. Therefore, a loss function is needed to bound the RRS factors. To this end, we design $\mathcal{L}_{\text{rrs}}$ as shown in \Beqref{eq:loss-function_rrs}. Here, $N_{\text{samples}}$ denotes the number of training samples in the current frame.
\begin{equation}
	\mathcal{L}_{\text{rrs}} = \sum_{i=1}^{N_{\text{samples}}} \left({q_{\text{nrrs}}(\theta) - q_{\text{norm}}(\theta)}\right)^2.
	\label{eq:loss-function_rrs}
\end{equation}
Note that $\mathcal{L}_{\text{rrs}}$ is defined on the sample points, similar to StatNet. Here, $q_{\text{norm}}(\theta)$ refers to the RRS factors after normalization (see \Bsecref{sec:methods_rrs-framework}). $\mathcal{L}_{\text{rrs}}$ can also help alleviate the issue that the behavior of RRS factors changes after normalization, as discussed in \Bsecref{sec:methods_rrs-framework}.

As for $\mathcal{L}_{\text{min}}$ and $\mathcal{L}_{\text{avg}}$, the pixel loss is equally shared among its samples. That is, if there are $k_{i}$ training samples for pixel $i$, then each training sample receives $\frac{1}{k_{i}}$ of the original loss.

\subsection{Mix-Depth}
\label{sec:methods_mix-depth}
Network inference introduces additional overhead, so the results of NRRS may not always be optimal. In some cases, the efficiency gains from NRRS are insufficient to compensate for this overhead, making it advantageous to disable NRRS at certain bounces. Inspired by traditional heuristic strategies, we propose a new mechanism called Mix-Depth to further enhance the performance of NRRS.

In some traditional RR methods, RR is heuristically disabled at the first few bounces~\cite{mitsuba}; in other words, these approaches combine fixed RR (1.0) with their RR strategies. Inspired by this, we can also integrate NRRS with other RRS methods. With the help of StatNet, we obtain a neural version of ADRRS (denoted as ADRRS (NN)), which allows us to combine NRRS with ADRRS (NN) as well.

The key idea of Mix-Depth is to select a specific RRS method at each bounce. In our implementation, we enhance NRRS with ADRRS (NN) and fixed RR. To achieve this, we propose two algorithms: Brute-Force Search and Heuristic Search.

\paragraph{Brute-Force Search}
This approach is suitable for small maximum depths. It tries all possible combinations of NRRS, ADRRS (NN), and fixed RR at each bounce, and selects the best one based on efficiency estimation from one-sample-per-pixel result. The time complexity of this algorithm is $O(3^B)$, where $B$ is the maximum path depth. For UPT with a max depth of 6, the number of combinations is 729, making this method feasible. However, for a max depth of 10, the number of combinations increases to 59,049, which is too large to search exhaustively.

\paragraph{Heuristic Search}
This approach is more practical for large maximum depths. We segment the maximum depth into several parts, divided by a parameter $T_{\text{d}}$. For each segment, starting from the shallower bounces, we search for the best combination using Brute-Force Search and set the earlier bounces to the selected combination. The remaining bounces are set to fixed RR.

For example, if $T_{\text{d}}=6$ and the maximum depth is $10$, we first set the latter bounces $[7,10]$ to fixed RR, and then perform Brute-Force Search on bounces $[1,6]$. After finding the best combination, we fix bounces $[1,6]$ to the selected combination and then perform Brute-Force Search on bounces $(6,10]$. The number of combinations to be tested drops to $3^6 + 3^4 = 810$, which is much more manageable.

Note that Mix-Depth is a post-processing step and does not affect the training process of NRRS. It can be used to combine multiple RRS methods without additional overhead, provided they share the same cache structure.

\section{Implementation}
\label{sec:implementation}
\paragraph{System}
We implement our algorithm using Kiraray~\cite{NPM23}, a GPU renderer based on CUDA and Optix~\cite{OptiX}. The neural network components of NRRS and AID-NRRS are implemented using tiny-cuda-nn~\cite{TCNN}.

\paragraph{Network Architecture}
The hidden layers for StatNet, and RRSNet share the same architecture. They are all MLP with 3 linear layers of 32 neurons, with Leaky ReLU activation~\cite{leakyReLU15} after each layer.

\paragraph{Input Encoding}
StatNet uses three input features (see \Bsecref{sec:methods_network-structure}). The position $\p \in \mathbb{R}^3$ is encoded with a trainable multiresolution hash grid~\cite{ngp22} $G$ with 8 levels, and each level has 2 features. The coarsest level resolution of $G$ is 16, and the resolution doubles at each finer level. The outgoing direction $\wo \in \mathbb{R}^2$ (spherical coordinates) is encoded using one-blob encoding~\cite{NIS19} with 4 bins. The roughness $\alpha_{\p}\in \mathbb{R}$ is encoded with one-blob encoding with 8 bins after applying the remapping function $1-\mathrm{e}^{-\alpha_{\p}}$~\cite{NRC21}.

For RRSNet, NRRS has five input features (see \Bfigref{fig:nrrs}). The high dynamic range features $\estExpected{\est{\Lo(\x,\wo)}} \in \mathbb{R}^3$, $\estExpected{\est{\Lo^2(\x,\wo)}} \in \mathbb{R}^3$, $t_{\x} \in \mathbb{R}^3$, and $I_{\text{pixel}} \in \mathbb{R}$ (mean value) are encoded with the box-cox transform~\cite{boxcox18} with $\lambda=0.5$. The roughness $\alpha_{\p} \in \mathbb{R}$ is encoded by the remapping function $1-\mathrm{e}^{-\alpha_{\p}}$.

AID-NRRS also have five input features (see \Bfigref{fig:aid-nrrs}). The encodings of $\p \in \mathbb{R}^{3}$ and $\wo \in \mathbb{R}^2$ are the same as in StatNet. The encodings of $t_{\x} \in \mathbb{R}^3$ and $I_{\text{pixel}} \in \mathbb{R}$ are the same as NRRS. The roughness $\alpha_{\p} \in \mathbb{R}$ is encoded with one-blob encoding with 4 bins after applying the remapping function $1-\mathrm{e}^{-\alpha_{\p}}$.

\paragraph{Output \& Activation}
The output of StatNet is $\Lo \in \mathbb{R}^3$ and $\Lo^2 \in \mathbb{R}^3$. In our experiments, StatNet works well without any activation functions on their outputs. For RRSNet, the output is the RRS factor $q \in \mathbb{R}$. We use the activation function defined in \Beqref{eq:rrsnet_activation} to ensure the output is positive. This function is a modified version of softplus~\cite{softplus00} to improve computational stability.

\begin{equation}
	y = \begin{cases}
		\ln(1+\exp(x)) & \text{if } x < 0    \\
		0.5x+\ln2      & \text{if } x \ge 0. \\
	\end{cases}
	\label{eq:rrsnet_activation}
\end{equation}

\paragraph{Training}
Since RRSNet requires the output of StatNet to compute gradients, we train StatNet first. The training process is separated into two steps:
\begin{enumerate}[leftmargin=*]
	\item Train StatNet with the ground truth $\est{\Lo}$ and $\est{\Lo^2}$ from UPT. At the same time, train RRSNet to output $1.0$ using the relative MSE loss $\mathcal{L}_2$.
	\item Continue training StatNet as in step (1). For RRSNet, use the loss described in \Bsecref{sec:methods_loss-design}. First, calculate the variance of the current frame for each pixel and use this to compute the loss $\LossRRSNet$. Then, calculate the derivative of $\LossRRSNet$ with respect to the variance. Using the Monte Carlo estimation from \Beqref{eq:gradient_transfer_rr_mc} and \Beqref{eq:gradient_transfer_splitting_mc}, propagate the gradient to the output of RRSNet. The statistics required for \Beqref{eq:gradient_transfer_rr_mc} and \Beqref{eq:gradient_transfer_splitting_mc} are provided by StatNet.
\end{enumerate}

Note that the gradient of RRSNet does not flow to StatNet; they are trained separately. We train StatNet and RRSNet using the Adam optimizer~\cite{Adam17} with fixed learning rates of $0.005$ and $0.0003$, respectively, and a batch size of $2^{19}$. During inference, we use exponential moving average (EMA) as~\cite{NRC21} to reduce noise from Monte Carlo estimation. For $\LossRRSNet$, we use $\gamma_1=0.01$, $\gamma_2=0.05$, and $\gamma_3=0.01$ for all scenes. We perform grid search on Bathroom scenes (not included in the test set), and use these hyperparameters for all other scenes. See more details of grid search in the supplementary material.

\paragraph{Compromise}
During the training process, we do not have the ground truth $\Ipx$ for each pixel, so we use the accumulated result from the previous frame as the ground truth. To support infinite depth in UPT, we cannot normalize the RRS factors (\Bsecref{sec:methods_rrs-framework}) at every depth. Therefore, if the normalization factor $F_{\text{norm}} \ge 1$ (\Beqref{eq:F_norm}), we do not perform normalization. In other words, we only \textit{downscale} the RRS factors.

\paragraph{Tackle Low Spp}
During the training process of RRSNet in step (2), we need to calculate the variance of the current frame for each pixel. However, with only 1 sample per pixel (spp), the image is too noisy and make the training process unstable. To address this, we employ both temporal and spatial denoising. First, we use $\Ipx^{\text{acc}}$ instead of the current frame $\Ipx^{\text{cur}}$ to calculate the error $\mathcal{E}^{\prime}_{\text{px}}$, where $\Ipx^{\text{acc}}$ is updated as:
\begin{equation}
	\Ipx^{\text{acc}} = (1-\alpha_{\text{mix}}) \Ipx^{\text{acc}} + \alpha_{\text{mix}}\Ipx^{\text{cur}}.
	\label{eq:error_temporal}
\end{equation} We set $\alpha_{\text{mix}}=0.5$ in our implementation.
Next, we use the Optix denoiser~\cite{OptixDenoiser} to denoise $\mathcal{E}^{\prime}_{\text{px}}$ and obtain $\mathcal{E}_{\text{px}}$ for use in \Beqref{eq:loss-function_rrsnet}. The denoiser is provided with the normal buffer but no albedo buffer, as including the albedo buffer can cause the error to resemble albedo rather than variance. Denoising introduces bias into the error estimation; therefore, we disable the Optix denoiser during the final 5 seconds of training.

\begin{figure*}[htbp]
	\centering
	\includegraphics[width=\linewidth]{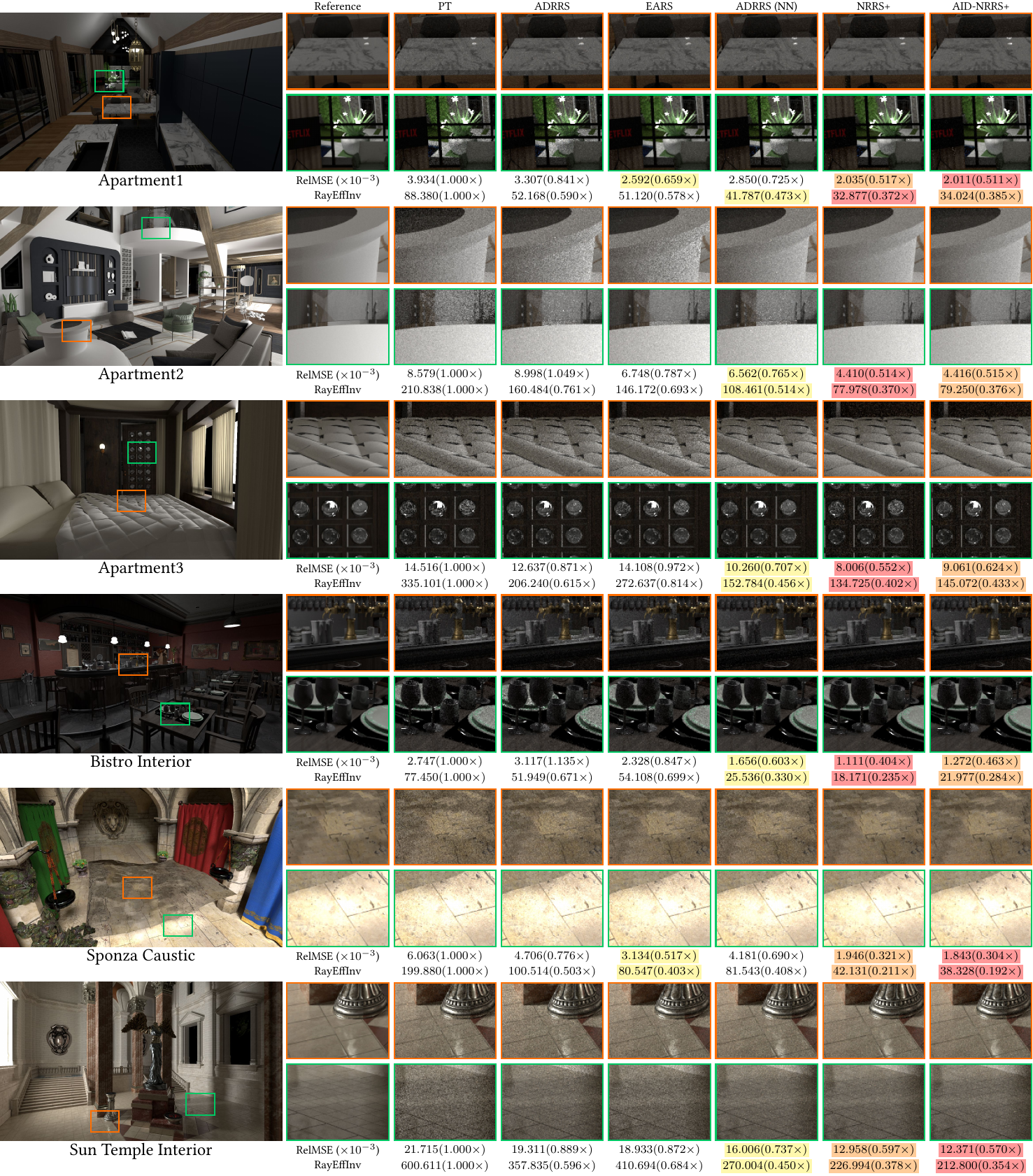}
	\caption{\textbf{Methods Evaluation and Comparisons on diverse scenes.}
		Each scene is trained for 60 seconds and then inferred for 60 seconds across all methods. All images are rendered at a resolution of $1280\times720$. Procedural results generated during training are not included due to excessive noise. The first row shows the RelMSE error ($\times 10^{-3}$) \textbf{over the entire image}, where lower values indicate better performance. The second row presents RayEffInv, as defined in \Bsecref{sec:results_comparisons}. Ray count is calculated as the total number of rays traced during the 60-seconds inference period, divided by the number of pixels. We color code the {\fboxsep1pt\colorbox[RGB]{255, 153, 153}{first}}, {\fboxsep1pt\colorbox[RGB]{255, 204, 153}{second}}, and {\fboxsep1pt\colorbox[RGB]{255, 248, 173}{third}} lowest numbers. Our approach consistently shows the best results using fewer rays, indicating that each traced ray was utilized as effectively as possible.
	}
	\label{fig:exp-main}
\end{figure*}

\section{Results and Discussion}
\label{sec:results}
All experiments are conducted on an Intel Core i7-12700F CPU and an NVIDIA RTX 3080 GPU. For the baseline settings, we enable next event estimation (NEE) and set the maximum path length to 6 by default, unless otherwise stated, as the majority of the contribution comes from the first few bounces. For a fair comparison, reflectance factorization is not used in our experiments, as it is not employed in the compared methods.

All images are rendered at a resolution of 1280$\times$720, and image quality is evaluated using mean relative squared error (RelMSE)~\cite{relativeL2}.

\subsection{Comparisons}
\label{sec:results_comparisons}
We compare our method with ADRRS~\cite{ADRRS16} and EARS~\cite{EARS22}, evaluating both the original tree-based implementations and their neural network variants. We re-implement these methods in Kiraray~\cite{NPM23}, following the open-source code of EARS~\cite{EARSCODE}, and use settings as close as possible to those in the original papers.
In this section, a "+" suffix appended to each method denotes the incorporation of the Mix-Depth module described in \Bsecref{sec:methods_mix-depth}. For instance, NRRS+ denotes NRRS with Mix-Depth applied.

\subsubsection{Equal-time Comparison}
\label{sec:results_equal_time}
For all methods, we train for 60 seconds and render for 60 seconds. Note that for NRRS+, the training time includes both the training of StatNet and RRSNet, as well as the time spent searching for the best strategy. Please refer to the supplementary material for a detailed breakdown of the training time distribution.

The comparison is performed across a variety of scenes. Apartment1 is a night scene illuminated by several small light sources. Apartment2 is a complex indoor environment containing many objects and lit by an area light positioned outside. Apartment3 features intricate indoor geometry and is illuminated by an area light outside. Bistro Interior is an indoor scene with diverse materials and numerous objects, illuminated by small light sources. Sponza Caustic is an outdoor scene exhibiting caustics produced by wave surfaces above the floor, and is illuminated by an area light. Sun Temple Interior contains a variety of glossy and metallic materials and is illuminated by an area light on the left. We also provide more scenes in the supplementary material.

\begin{figure*}[htbp]
	\centering
	\includegraphics[width=\textwidth]{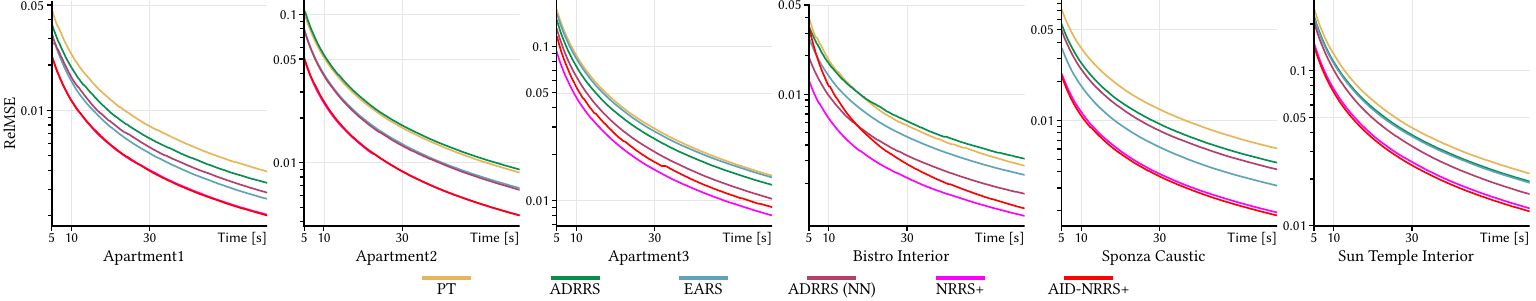}
	\caption{
		RelMSE convergence curves for different methods corresponding to \Bfigref{fig:exp-main}.
	}
	\label{fig:exp-main-curve}
\end{figure*}

The results are presented in \Bfigref{fig:exp-main} and the upper half of \Btabref{tab:ab-mix-depth}. It can be observed that NRRS+ and AID-NRRS+ consistently outperform the other methods. With the help of neural networks, ADRRS also shows improvement, but still does not match the performance of our approach. \Bfigref{fig:exp-main-curve} shows the corresponding convergence curves, illustrating how RelMSE evolves over time. We also provide error maps in \Bfigref{fig:exp-main-errors-ears-vs-nrrs+} (see supplementary material for more), which highlight the differences in error distribution between EARS and NRRS+.

\subsubsection{Ray Efficiency Comparison}
Here, we compare the ray efficiency of different methods. We define the metric \textit{RayEff} as the reciprocal of the product of the number of rays and RelMSE within the same rendering time, and use its inverse, \textit{RayEffInv}, for evaluation. This is similar to the equal spp comparison in path guiding~\cite{PPG17,NPM23}, as we ignore the time spent on generating rays for all methods. The results are shown in \Bfigref{fig:exp-main}, using the same experimental settings as in the \textit{Equal Time Comparison}. The results demonstrate that our method can achieve the same RelMSE with fewer rays.

\subsubsection{Deeper Depth}
\label{sec:results_deeper_depth}
We also explore the performance of our method with deeper depth settings. We set the maximum path length to 10 and perform the same experiments as \Bsecref{sec:results_equal_time}. The results are shown in \Btabref{tab:main-d10} and the lower half of \Btabref{tab:ab-mix-depth}. It can be observed that NRRS+ and AID-NRRS+ still outperform the other methods. Here we only show the numerical results, the detailed images can be found in the supplementary material.

\begin{figure}[htbp]
	\centering
	\includegraphics[width=\linewidth]{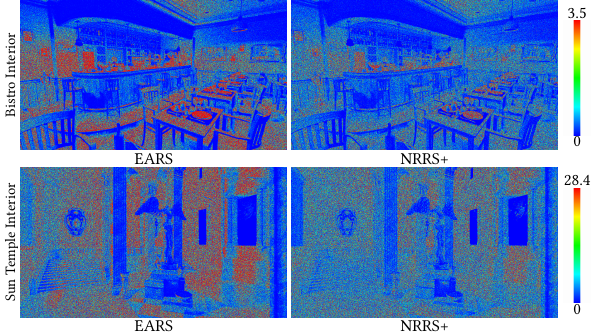}
	\caption{\textbf{EARS Vs. NRRS+ in terms of RelMSE ($\times 10^{-3}$).}
		For each scene, the maximum tint value (corresponding to red) is set to 1.5 times the largest RelMSE across all methods.
	}
	\label{fig:exp-main-errors-ears-vs-nrrs+}
\end{figure}

\begin{table*}[htbp]
	\centering
	\caption{\textbf{Main Experiment for Depth 10.} The experiment setting are the same with \Bfigref{fig:exp-main}, except the maximum path depth is set to 10. The data in the table is RelMSE ($\times 10^{-3}$). We also color code the {\fboxsep1pt\colorbox[RGB]{255, 153, 153}{first}}, {\fboxsep1pt\colorbox[RGB]{255, 204, 153}{second}}, and {\fboxsep1pt\colorbox[RGB]{255, 248, 173}{third}} lowest numbers.}
	\label{tab:main-d10}
	\begin{adjustbox}{max width=\linewidth}
		\begin{tabular}{ccccccc}
			\toprule
			Scene               & PT                                                 & ADRRS                                              & EARS                                                         & ADRRS (NN)                                                    & NRRS+                                                         & AID-NRRS+                                                     \\
			\midrule
			Apartment1          & \fboxsep1pt\colorbox{white}{5.600(1.000$\times$)}  & \fboxsep1pt\colorbox{white}{4.563(0.815$\times$)}  & \fboxsep1pt\colorbox[RGB]{255,248,173}{3.245(0.580$\times$)} & \fboxsep1pt\colorbox{white}{3.345(0.597$\times$)}             & \fboxsep1pt\colorbox[RGB]{255,204,153}{2.692(0.481$\times$)}  & \fboxsep1pt\colorbox[RGB]{255,153,153}{2.368(0.423$\times$)}  \\
			Apartment2          & \fboxsep1pt\colorbox{white}{12.376(1.000$\times$)} & \fboxsep1pt\colorbox{white}{9.363(0.757$\times$)}  & \fboxsep1pt\colorbox{white}{9.270(0.749$\times$)}            & \fboxsep1pt\colorbox[RGB]{255,248,173}{6.962(0.563$\times$)}  & \fboxsep1pt\colorbox[RGB]{255,153,153}{5.769(0.466$\times$)}  & \fboxsep1pt\colorbox[RGB]{255,204,153}{5.805(0.469$\times$)}  \\
			Apartment3          & \fboxsep1pt\colorbox{white}{20.855(1.000$\times$)} & \fboxsep1pt\colorbox{white}{14.659(0.703$\times$)} & \fboxsep1pt\colorbox{white}{21.036(1.009$\times$)}           & \fboxsep1pt\colorbox[RGB]{255,248,173}{11.555(0.554$\times$)} & \fboxsep1pt\colorbox[RGB]{255,153,153}{8.940(0.429$\times$)}  & \fboxsep1pt\colorbox[RGB]{255,204,153}{9.332(0.447$\times$)}  \\
			Bistro Interior     & \fboxsep1pt\colorbox{white}{4.488(1.000$\times$)}  & \fboxsep1pt\colorbox{white}{3.832(0.854$\times$)}  & \fboxsep1pt\colorbox{white}{3.099(0.691$\times$)}            & \fboxsep1pt\colorbox[RGB]{255,248,173}{2.040(0.455$\times$)}  & \fboxsep1pt\colorbox[RGB]{255,204,153}{1.303(0.290$\times$)}  & \fboxsep1pt\colorbox[RGB]{255,153,153}{1.299(0.289$\times$)}  \\
			Sponza Caustic      & \fboxsep1pt\colorbox{white}{8.782(1.000$\times$)}  & \fboxsep1pt\colorbox{white}{5.271(0.600$\times$)}  & \fboxsep1pt\colorbox[RGB]{255,248,173}{3.544(0.403$\times$)} & \fboxsep1pt\colorbox{white}{4.568(0.520$\times$)}             & \fboxsep1pt\colorbox[RGB]{255,204,153}{2.172(0.247$\times$)}  & \fboxsep1pt\colorbox[RGB]{255,153,153}{2.146(0.244$\times$)}  \\
			Sun Temple Interior & \fboxsep1pt\colorbox{white}{35.431(1.000$\times$)} & \fboxsep1pt\colorbox{white}{24.871(0.702$\times$)} & \fboxsep1pt\colorbox{white}{27.760(0.783$\times$)}           & \fboxsep1pt\colorbox[RGB]{255,248,173}{18.485(0.522$\times$)} & \fboxsep1pt\colorbox[RGB]{255,204,153}{16.061(0.453$\times$)} & \fboxsep1pt\colorbox[RGB]{255,153,153}{15.592(0.440$\times$)} \\
			\bottomrule
		\end{tabular}
	\end{adjustbox}
\end{table*}

\subsection{Ablations}
\label{sec:results_ablation}

In this section, we conduct ablation studies to evaluate the effectiveness of essential components in our method, including the Mix-Depth module, Denoise module, RRSNet and loss design.

\subsubsection{Mix-Depth Module}
\label{sec:results_ablation_mix_depth}
We perform Mix-Depth ablation studies on NRRS, AID-NRRS, and EARS. EARS+ denotes a combination of EARS, ADRRS, and fixed RR. Since EARS and ADRRS share the same cache structure, ADRRS can be incorporated into EARS. The results are presented in \Btabref{tab:ab-mix-depth}. We can see they are all improved by the Mix-Depth module in most scenes. For EARS, the reasons are twofold: first, certain approximations are made in the implementation for practical purposes, such as tree-structured caching; second, under the normalization framework, EARS is no longer the optimal solution.

However, Mix-Depth does not always lead to performance improvements. For instance, in the Sponza Caustic scene with maximum path length set to 10, EARS+ performs slightly worse than EARS. This is because efficiency estimation based on one sample per pixel is inherently noisy, and Heuristic Search can only find a locally optimal strategy as the search space is large.

\begin{table*}[htbp]
	\centering
	\caption{\textbf{Ablation study for Mix-Depth Module.} The experiment setting are the same with \Bfigref{fig:exp-main}, except the maximum path depth for the scenes in the \textbf{upper half} of the table is \textbf{6}, while for those in the \textbf{lower half} it is \textbf{10}. The data in the table is RelMSE ($\times 10^{-3}$). We also color code the {\fboxsep1pt\colorbox[RGB]{255, 153, 153}{first}}, {\fboxsep1pt\colorbox[RGB]{255, 204, 153}{second}}, and {\fboxsep1pt\colorbox[RGB]{255, 248, 173}{third}} lowest numbers.}
	\label{tab:ab-mix-depth}
	\begin{adjustbox}{max width=\linewidth}
		\begin{tabular}{cccccccc}
			\toprule
			Scene               & PT                                                 & NRRS                                                          & NRRS+                                                         & AID-NRRS                                                      & AID-NRRS+                                                     & EARS                                                         & EARS+                                                        \\
			\midrule
			Apartment1          & \fboxsep1pt\colorbox{white}{3.934(1.000$\times$)}  & \fboxsep1pt\colorbox[RGB]{255,248,173}{2.344(0.596$\times$)}  & \fboxsep1pt\colorbox[RGB]{255,204,153}{2.035(0.517$\times$)}  & \fboxsep1pt\colorbox{white}{2.364(0.601$\times$)}             & \fboxsep1pt\colorbox[RGB]{255,153,153}{2.011(0.511$\times$)}  & \fboxsep1pt\colorbox{white}{2.592(0.659$\times$)}            & \fboxsep1pt\colorbox{white}{2.580(0.656$\times$)}            \\
			Apartment2          & \fboxsep1pt\colorbox{white}{8.579(1.000$\times$)}  & \fboxsep1pt\colorbox{white}{5.241(0.611$\times$)}             & \fboxsep1pt\colorbox[RGB]{255,153,153}{4.410(0.514$\times$)}  & \fboxsep1pt\colorbox[RGB]{255,248,173}{5.203(0.607$\times$)}  & \fboxsep1pt\colorbox[RGB]{255,204,153}{4.416(0.515$\times$)}  & \fboxsep1pt\colorbox{white}{6.748(0.787$\times$)}            & \fboxsep1pt\colorbox{white}{5.474(0.638$\times$)}            \\
			Apartment3          & \fboxsep1pt\colorbox{white}{14.516(1.000$\times$)} & \fboxsep1pt\colorbox[RGB]{255,204,153}{8.852(0.610$\times$)}  & \fboxsep1pt\colorbox[RGB]{255,153,153}{8.006(0.552$\times$)}  & \fboxsep1pt\colorbox{white}{9.429(0.650$\times$)}             & \fboxsep1pt\colorbox[RGB]{255,248,173}{9.061(0.624$\times$)}  & \fboxsep1pt\colorbox{white}{14.108(0.972$\times$)}           & \fboxsep1pt\colorbox{white}{11.808(0.813$\times$)}           \\
			Bistro Interior     & \fboxsep1pt\colorbox{white}{2.747(1.000$\times$)}  & \fboxsep1pt\colorbox{white}{1.955(0.712$\times$)}             & \fboxsep1pt\colorbox[RGB]{255,153,153}{1.111(0.404$\times$)}  & \fboxsep1pt\colorbox[RGB]{255,248,173}{1.635(0.595$\times$)}  & \fboxsep1pt\colorbox[RGB]{255,204,153}{1.272(0.463$\times$)}  & \fboxsep1pt\colorbox{white}{2.328(0.847$\times$)}            & \fboxsep1pt\colorbox{white}{2.325(0.846$\times$)}            \\
			Sponza Caustic      & \fboxsep1pt\colorbox{white}{6.063(1.000$\times$)}  & \fboxsep1pt\colorbox[RGB]{255,248,173}{2.821(0.465$\times$)}  & \fboxsep1pt\colorbox[RGB]{255,204,153}{1.946(0.321$\times$)}  & \fboxsep1pt\colorbox{white}{2.866(0.473$\times$)}             & \fboxsep1pt\colorbox[RGB]{255,153,153}{1.843(0.304$\times$)}  & \fboxsep1pt\colorbox{white}{3.134(0.517$\times$)}            & \fboxsep1pt\colorbox{white}{3.084(0.509$\times$)}            \\
			Sun Temple Interior & \fboxsep1pt\colorbox{white}{21.715(1.000$\times$)} & \fboxsep1pt\colorbox{white}{14.925(0.687$\times$)}            & \fboxsep1pt\colorbox[RGB]{255,204,153}{12.958(0.597$\times$)} & \fboxsep1pt\colorbox[RGB]{255,248,173}{14.860(0.684$\times$)} & \fboxsep1pt\colorbox[RGB]{255,153,153}{12.371(0.570$\times$)} & \fboxsep1pt\colorbox{white}{18.933(0.872$\times$)}           & \fboxsep1pt\colorbox{white}{17.126(0.789$\times$)}           \\
			\midrule
			Apartment1          & \fboxsep1pt\colorbox{white}{5.600(1.000$\times$)}  & \fboxsep1pt\colorbox{white}{3.387(0.605$\times$)}             & \fboxsep1pt\colorbox[RGB]{255,204,153}{2.692(0.481$\times$)}  & \fboxsep1pt\colorbox[RGB]{255,248,173}{3.161(0.565$\times$)}  & \fboxsep1pt\colorbox[RGB]{255,153,153}{2.368(0.423$\times$)}  & \fboxsep1pt\colorbox{white}{3.245(0.580$\times$)}            & \fboxsep1pt\colorbox{white}{3.205(0.572$\times$)}            \\
			Apartment2          & \fboxsep1pt\colorbox{white}{12.376(1.000$\times$)} & \fboxsep1pt\colorbox{white}{7.892(0.638$\times$)}             & \fboxsep1pt\colorbox[RGB]{255,153,153}{5.769(0.466$\times$)}  & \fboxsep1pt\colorbox{white}{7.899(0.638$\times$)}             & \fboxsep1pt\colorbox[RGB]{255,204,153}{5.805(0.469$\times$)}  & \fboxsep1pt\colorbox{white}{9.270(0.749$\times$)}            & \fboxsep1pt\colorbox[RGB]{255,248,173}{7.496(0.606$\times$)} \\
			Apartment3          & \fboxsep1pt\colorbox{white}{20.855(1.000$\times$)} & \fboxsep1pt\colorbox{white}{14.632(0.702$\times$)}            & \fboxsep1pt\colorbox[RGB]{255,153,153}{8.940(0.429$\times$)}  & \fboxsep1pt\colorbox[RGB]{255,248,173}{14.474(0.694$\times$)} & \fboxsep1pt\colorbox[RGB]{255,204,153}{9.332(0.447$\times$)}  & \fboxsep1pt\colorbox{white}{21.036(1.009$\times$)}           & \fboxsep1pt\colorbox{white}{17.832(0.855$\times$)}           \\
			Bistro Interior     & \fboxsep1pt\colorbox{white}{4.488(1.000$\times$)}  & \fboxsep1pt\colorbox{white}{2.762(0.615$\times$)}             & \fboxsep1pt\colorbox[RGB]{255,204,153}{1.303(0.290$\times$)}  & \fboxsep1pt\colorbox[RGB]{255,248,173}{2.261(0.504$\times$)}  & \fboxsep1pt\colorbox[RGB]{255,153,153}{1.299(0.289$\times$)}  & \fboxsep1pt\colorbox{white}{3.099(0.691$\times$)}            & \fboxsep1pt\colorbox{white}{2.474(0.551$\times$)}            \\
			Sponza Caustic      & \fboxsep1pt\colorbox{white}{8.782(1.000$\times$)}  & \fboxsep1pt\colorbox{white}{4.341(0.494$\times$)}             & \fboxsep1pt\colorbox[RGB]{255,204,153}{2.172(0.247$\times$)}  & \fboxsep1pt\colorbox{white}{4.964(0.565$\times$)}             & \fboxsep1pt\colorbox[RGB]{255,153,153}{2.146(0.244$\times$)}  & \fboxsep1pt\colorbox[RGB]{255,248,173}{3.544(0.403$\times$)} & \fboxsep1pt\colorbox{white}{4.222(0.481$\times$)}            \\
			Sun Temple Interior & \fboxsep1pt\colorbox{white}{35.431(1.000$\times$)} & \fboxsep1pt\colorbox[RGB]{255,248,173}{23.690(0.669$\times$)} & \fboxsep1pt\colorbox[RGB]{255,204,153}{16.061(0.453$\times$)} & \fboxsep1pt\colorbox{white}{25.792(0.728$\times$)}            & \fboxsep1pt\colorbox[RGB]{255,153,153}{15.592(0.440$\times$)} & \fboxsep1pt\colorbox{white}{27.760(0.783$\times$)}           & \fboxsep1pt\colorbox{white}{25.459(0.719$\times$)}           \\
			\bottomrule
		\end{tabular}
	\end{adjustbox}
\end{table*}

\subsubsection{Denoise Module}
\label{sec:results_ablation_denoise}
\begin{figure}[htbp]
	\centering
	\includegraphics[width=\linewidth]{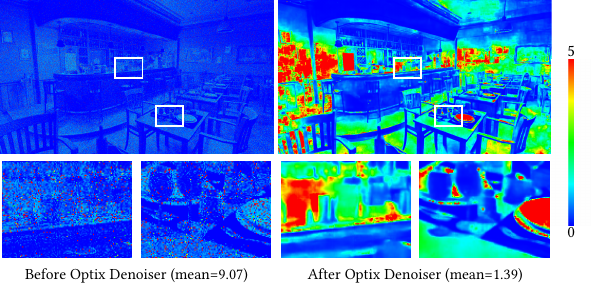}
	\caption{\textbf{Training Errors visualization before/after applying Optix denoiser.}
		The average error over the entire image is included.
	}
	\label{fig:exp-error-before-and-after-denoising}
\end{figure}

The denoise module is introduced to address instability during training at low samples per pixel (see \Bsecref{sec:implementation}). Here, we focus on spatial denoising using the OptiX denoiser. \Bfigref{fig:exp-error-before-and-after-denoising} shows the error before and after applying the OptiX denoiser. It can be seen that the OptiX denoiser effectively reduces noise while preserving the main structure, and also decreases the absolute value of the error.

We conduct ablation studies on NRRS to evaluate its functionality. The results are shown in \Btabref{tab:ablations}. It can be observed that always using the denoiser leads to a significant drop in performance, as the denoiser introduces bias into the error estimation. Conversely, disabling the denoiser results in unstable training, which often leads to NaN values. Our strategy achieves superior results in most scenes, demonstrating its effectiveness. However, we have not yet found the optimal allocation strategy for dividing time between denoising and non-denoising phases; we leave this for future work. For example, in Sponza Caustic, the denoiser is not beneficial.

\subsubsection{RRSNet}
\label{sec:results_ablation_earsnn}
Simply replacing the tree structure in EARS with a neural network can not achieve better performance, as shown in \Btabref{tab:ablations}. We use two variants of EARS based on how the mean of the cost is handled: EARS (NN) predicts it jointly with StatNet (by adding a relative loss for the cost to \Beqref{eq:loss-function_StatNet}), while EARS (NT) uses an additional tree to cache it. EARS (NN) performs poorly, likely because it is challenging for a simple MLP to accurately predict all three values. Although EARS (NT) can provide better RRS factors, the additional overhead of querying the tree means it still does not outperform EARS, let alone NRRS. Additionally, neither method can address the normalization problem mentioned in \Bsecref{sec:methods_rrs-framework}.

\begin{table*}[htbp]
	\centering
	\caption{\textbf{Ablation study.}
	"Denoise Always" shows results when denoising is always enabled, while "No Denoise" shows results when denoising is always disabled. EARS (NN) and EARS (NT) are two variants of EARS: EARS (NN) uses a neural network to cache all statistics, while EARS (NT) uses a neural network to cache the mean and variance of radiance and a tree to cache the mean of the cost. "No $\mathcal{L}_{\text{avg}}$" corresponds to NRRS with $\mathcal{L}_{\text{avg}}$ disabled. Errors are reported as RelMSE ($\times 10^{-3}$). The best result for each row is \fboxsep1pt\colorbox[RGB]{255,153,153}{highlighted}.
	}
	\label{tab:ablations}
	\begin{tabular}{c|c|cc|ccc|c}
		\toprule
		                    & NRRS                                           & Denoise Always & No Denoise                                    & EARS   & EARS (NN) & EARS (NT) & no $\mathcal{L}_{\text{avg}}$ \\
		\midrule
		Apartment1          & \fboxsep1pt\colorbox[RGB]{255,153,153}{2.344}  & 4.049          & nan                                           & 2.592  & 4.330     & 3.545     & 2.448                         \\
		Apartment2          & \fboxsep1pt\colorbox[RGB]{255,153,153}{5.241}  & 8.945          & nan                                           & 6.748  & 7.348     & 6.543     & 5.315                         \\
		Apartment3          & \fboxsep1pt\colorbox[RGB]{255,153,153}{8.852}  & 15.008         & nan                                           & 14.108 & 17.280    & 14.477    & 9.199                         \\
		Bistro Interior     & \fboxsep1pt\colorbox[RGB]{255,153,153}{1.955}  & 3.142          & 2.264                                         & 2.328  & 2.951     & 2.770     & 2.112                         \\
		Sponza Caustic      & 2.821                                          & 6.338          & \fboxsep1pt\colorbox[RGB]{255,153,153}{2.737} & 3.134  & 5.457     & 4.841     & 2.921                         \\
		Sun Temple Interior & \fboxsep1pt\colorbox[RGB]{255,153,153}{14.925} & 23.039         & nan                                           & 18.933 & 23.287    & 23.562    & 14.948                        \\
		\bottomrule
	\end{tabular}
\end{table*}

\subsubsection{Loss Design}
\label{sec:results_ablation_loss-design}

\begin{figure}[htbp]
	\centering
	\includegraphics[width=\linewidth]{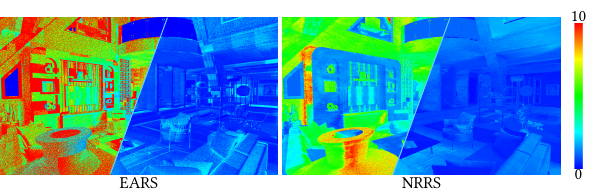}
	\caption{\textbf{Visualization of the RRS factors for EARS (left) and NRRS (right), respectively.} For each method, the upper-left region shows the original RRS factors $q_{\text{orig}}$ before applying the RRS module, and the lower-right region shows the factors $q_{\text{norm}}$ after it. Both are computed at the primary ray intersection point prior to the RRS stage (\Bfigref{fig:wavefront}).
	}
	\label{fig:rrs-before-and-after}
\end{figure}

There are three loss functions for RRSNet: $\mathcal{L}_{\text{min}}$, $\mathcal{L}_{\text{avg}}$, and $\mathcal{L}_{\text{rrs}}$. As mentioned in \Bsecref{sec:methods_loss-design}, $\mathcal{L}_{\text{min}}$ is necessary to reduce the error in the final image. $\mathcal{L}_{\text{rrs}}$ is used to bound the RRS factors and narrow the gap between pre- and post-normalization. \Bfigref{fig:rrs-before-and-after} shows that, although the problem is not perfectly solved, the gap between pre- and post-normalization for NRRS is smaller than for EARS. This demonstrates that $\mathcal{L}_{\text{rrs}}$ is helpful. Furthermore, we conduct an ablation study (See \Btabref{tab:ablations}) to show that $\mathcal{L}_{\text{avg}}$ can slightly improve performance.

\subsection{Evaluation \& Validation}
\label{sec:results_evaluation}
\subsubsection{Encoding of Position}
\label{sec:results_evaluation_encoding}

\begin{figure}[htbp]
	\centering
	\begin{subfigure}[b]{0.45\linewidth}
		\centering
		\includegraphics[width=\linewidth]{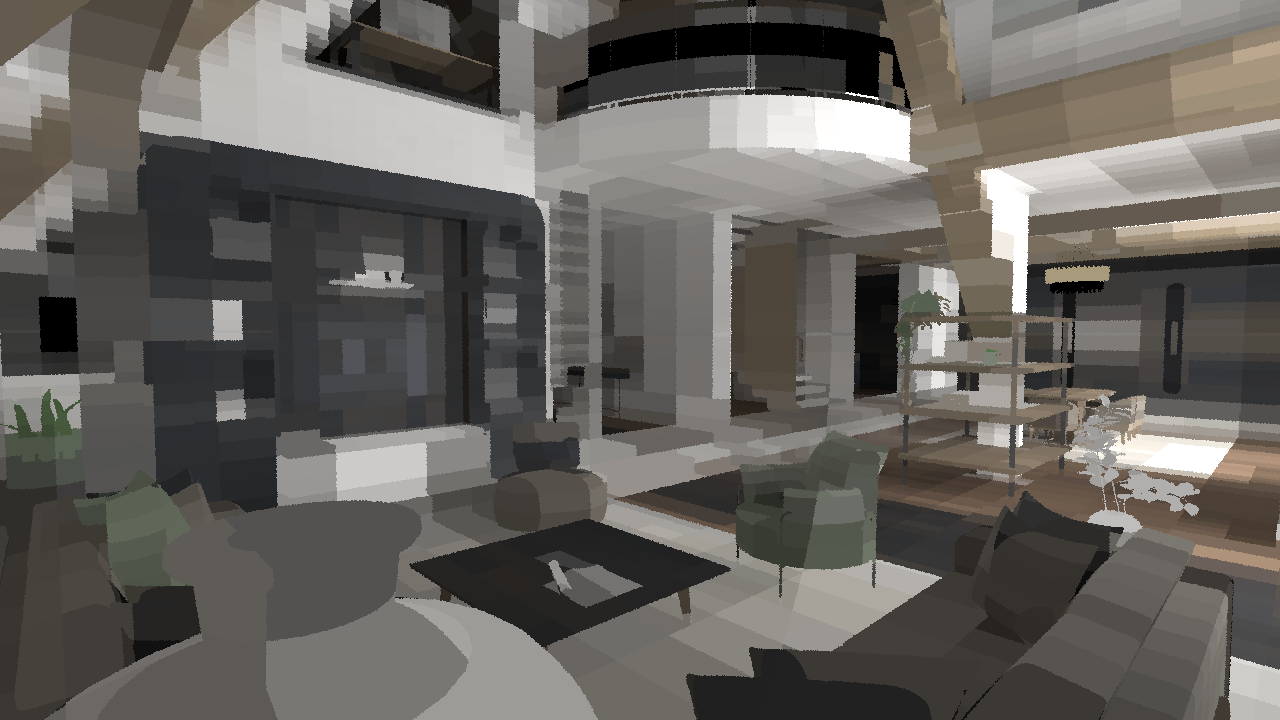}
		\caption{\textbf{Tree Cache}}
		\label{fig:sub-Lcache_tree}
	\end{subfigure}
	\hfill
	\begin{subfigure}[b]{0.45\linewidth}
		\centering
		\includegraphics[width=\linewidth]{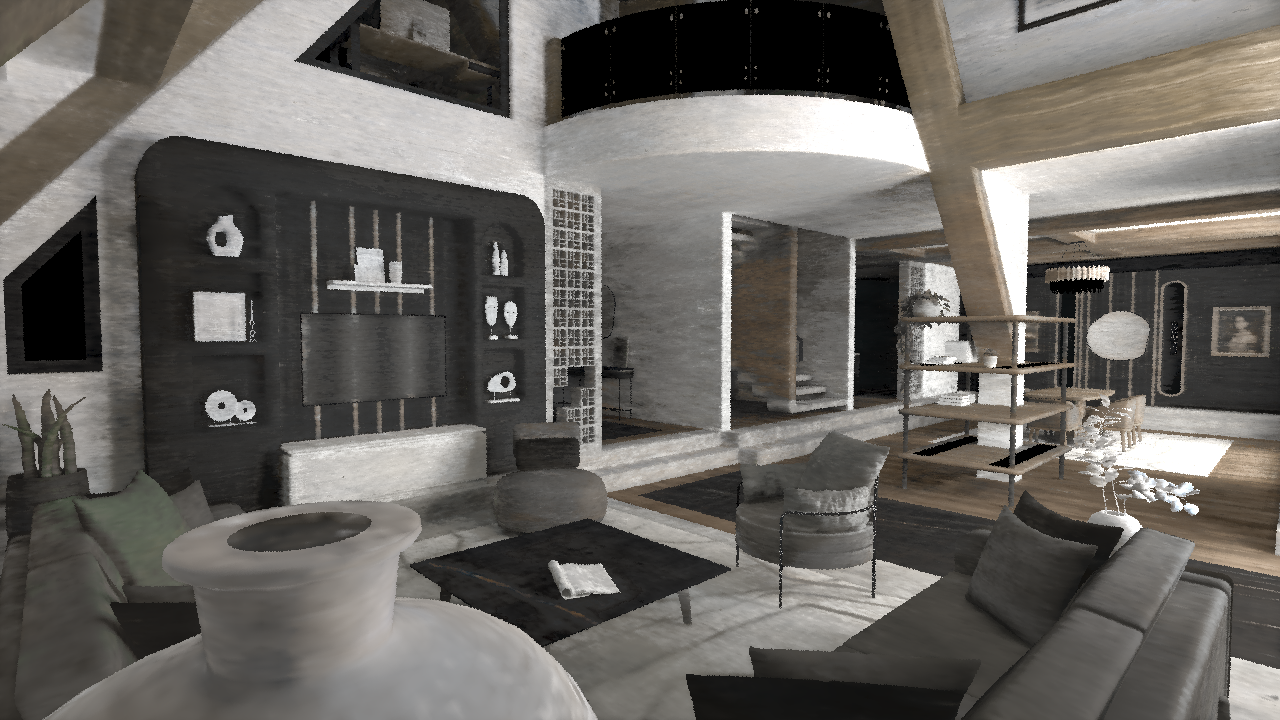}
		\caption{\textbf{Network Cache}}
		\label{fig:sub-Lcache_network}
	\end{subfigure}
	\caption{\textbf{Radiance Cache Visualization.}
		Visualization of the radiance cache from the tree structure and the neural network. The left shows the radiance cache from the tree structure, and the right shows the radiance cache from the network. Both caches are queried at the primary ray intersection point.
	}
	\label{fig:Lcache}
\end{figure}

In \Bfigref{fig:exp-main}, a vase near the camera plane in Apartment2 is visible. Traditional tree-based structures struggle to capture statistical variations in such a small region under limited memory, whereas our neural network with hash-grid encoding handles this efficiently. \Bfigref{fig:sub-Lcache_tree} and \Bfigref{fig:sub-Lcache_network} show the radiance cache from the tree structure and the neural network, respectively. The radiance from neural network is more accurate than that from the tree structure.

\subsubsection{StatNet}
\label{sec:results_evaluation_statnet}
StatNet outputs the mean and variance of radiance. In our implementation, both the encoding and all layers of StatNet are shared between these outputs. We design several sharing strategies and compare them in \Btabref{tab:ab-statnet}. The memory usage of each method is set to be approximately equal (see the supplementary material for details). The less the two outputs share, the less interference occurs between them; however, this increases both training and inference time. Each design represents a trade-off between accuracy and efficiency. For NRRS/NRRS+, StatNet is required during inference, and the results show that sharing all layers achieves the best performance. For AID-NRRS/AID-NRRS+, StatNet is not required during inference, so the trade-off only affects training. The results indicate that sharing all layers is not always the best choice in this case. We have not further explored which sharing strategy is optimal under specific time constraints for AID-NRRS/AID-NRRS+, leaving this for future work.

\begin{table}[htbp]
	\centering
	\caption{\textbf{Different design for StatNet.}
		StatNet outputs the mean and variance of radiance. "All" indicates that both the encoding and all layers of StatNet are shared; "PartNet" means the encoding and the first several layers are shared; "Encoding" means only the encoding is shared; and "None" means no sharing. Errors are reported as RelMSE ($\times 10^{-3}$). The best results are \fboxsep1pt\colorbox[RGB]{255,153,153}{highlighted} for each column.
	}
	\label{tab:ab-statnet}
	\begin{adjustbox}{max width=\linewidth}
		\begin{tabular}{ccccc}
			\toprule
			         & NRRS                                           & NRRS+                                          & AID-NRRS                                       & AID-NRRS+                                      \\
			\midrule
			\multicolumn{5}{l}{\textbf{Bistro Interior}}                                                                                                                                                                 \\
			All      & \fboxsep1pt\colorbox[RGB]{255,153,153}{1.955}  & \fboxsep1pt\colorbox[RGB]{255,153,153}{1.111}  & 1.635                                          & 1.272                                          \\
			PartNet  & 2.272                                          & 1.243                                          & \fboxsep1pt\colorbox[RGB]{255,153,153}{1.620}  & 1.214                                          \\
			Encoding & 2.183                                          & 1.133                                          & 1.957                                          & \fboxsep1pt\colorbox[RGB]{255,153,153}{1.105}  \\
			None     & 2.501                                          & 1.305                                          & 2.116                                          & 1.186                                          \\
			\midrule
			\multicolumn{5}{l}{\textbf{Sun Temple Interior}}                                                                                                                                                             \\
			All      & \fboxsep1pt\colorbox[RGB]{255,153,153}{14.925} & \fboxsep1pt\colorbox[RGB]{255,153,153}{12.958} & \fboxsep1pt\colorbox[RGB]{255,153,153}{14.860} & \fboxsep1pt\colorbox[RGB]{255,153,153}{12.371} \\
			PartNet  & 16.376                                         & 17.447                                         & 15.491                                         & 13.021                                         \\
			Encoding & 15.919                                         & 17.532                                         & 15.449                                         & 12.933                                         \\
			None     & 17.230                                         & 18.132                                         & 15.123                                         & 13.554                                         \\
			\bottomrule
		\end{tabular}
	\end{adjustbox}
\end{table}

\subsubsection{Caustics}
\label{sec:results_evaluation_caustics}
\begin{figure}[htbp]
	\centering
	\includegraphics[width=\linewidth]{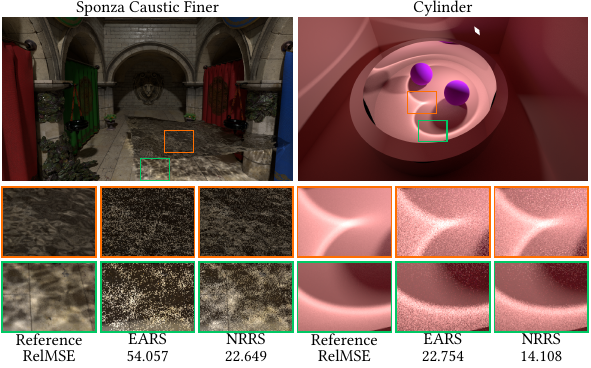}
	\caption{\textbf{Detailed Caustics Scene.}
		Comparison of different methods in two detailed caustics. The settings are the same as in \Bfigref{fig:exp-main}, and the RelMSE ($\times 10^{-3}$) is calculated \textbf{over the entire image}.
	}
	\label{fig:exp-caustics}
\end{figure}

We evaluate our method on more challenging scenes featuring detailed caustics; the results are shown in \Bfigref{fig:exp-caustics}. Sponza Caustic Finer is a more complex version of Sponza Caustic, with higher precision in the wave surface and a smaller area light. Cylinder is a scene with a mirror cylinder above a plane, illuminated by an area light. The results demonstrate that our method consistently outperforms EARS in these two challenging scenarios.

\subsubsection{Failure Cases}
\begin{figure*}[htbp]
	\centering
	\includegraphics[width=\textwidth]{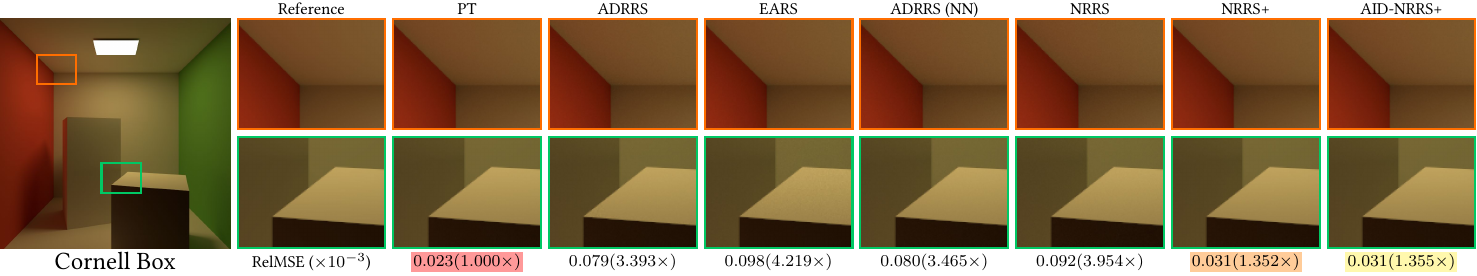}
	\caption{\textbf{Failure Case.}
		Cornell Box is a simple scene with a light source. The experimental setting is the same as in \Bfigref{fig:exp-main}, except the resolution is $750\times750$. Under such simple geometry and illumination conditions, the original path tracing outperforms the other methods.
	}
	\label{fig:exp-failure-case}
\end{figure*}

\Bfigref{fig:exp-failure-case} shows the result for the Cornell Box scene under simple lighting conditions. It can render approximately 160 frames per second with 1 ray per pixel in Kiraray. In this case, simple path tracing outperforms the other methods. With hardware acceleration, the overhead of the other techniques cannot be compensated. However, with our Mix-Depth design, NRRS+ can approximately fall back to simple path tracing.

\section{Conclusion, Limitation, and Future Work}
\label{sec:conclusion}

\paragraph{Conclusion}
In this paper, we propose a new framework for integrating RRS techniques into wavefront path tracing.
Since the theoretical foundations of traditional RRS methods do not hold under this framework, we introduce two neural RRS methods: NRRS and AID-NRRS. Both are carefully designed to address the challenges outlined above, with only subtle differences in their implementation.
Our Mix-Depth technique can also combine multiple RRS methods with little overhead. To our knowledge, we are the first to introduce neural networks into RRS techniques, and this surpasses the state-of-the-art RRS method~\cite{EARS22}. Our approach consistently shows the best results using fewer rays when compared to the alternative, indicating that each traced ray was utilized as effectively as possible in our approach.

\paragraph{Limitation}
NRRS leverages the relationship between pixel variance and the statistics of local radiance to optimize the RRS factors. However, these estimations are not always sufficiently accurate. Using only one sample per pixel (spp) is sometimes inadequate for estimating pixel variance, which can make the training process unstable. And our denoising technique only sacrifice the accuracy for the unstable. For local statistics, the variance estimation of StatNet is not always reliable, which can prevent the RRS factors from converging to the optimal result. Efficiently caching variance in the network remains challenging and is left for future work. Although our Mix-Depth technique is effective, finding the optimal strategy becomes difficult when there are many alternatives; even our heuristic method is impractical in such cases.

\paragraph{Future Work}
Our work offers several straightforward extensions. ADRRS~\cite{ADRRS16} can be further optimized by integrating path guiding with minimal overhead, as they share the same cache structure. NRRS can also be combined with other neural path guiding methods~\cite{NPM23,NPG-NASG23}, which can improve the quality of radiance samples and also enable joint optimization for better performance. \citet{MARS24} learns to distribute samples across multiple sampling techniques, which we have not explored; this could further improve the efficiency of RRS methods, especially since NEE can be ineffective in many scenarios.
As for Mix-Depth, a better mixing strategy would be to determine the RRS method at each scene intersection, rather than at each bounce.
Currently, NRRS queries the network at every bounce. If we could instead query the network in primary space, we would only need to query once per path, allowing for more complex network designs and potentially better performance. This is also an interesting direction for future research.

\bibliography{ref.bib}


\clearpage\newpage

\end{document}